\documentstyle[prb,aps,eqsecnum,floats]{revtex}

\begin{document} 
\draft
\twocolumn[
\hsize\textwidth\columnwidth\hsize\csname@twocolumnfalse\endcsname

\title{A unified description of static and dynamic properties of 
Fermi liquids  } 

\author{ N. Dupuis }
\address{Department of Physics, University of Maryland, College Park, MD
20742-4111, USA \\
Laboratoire de Physique des Solides, Universit\'e Paris-Sud, 91405
Orsay, France \cite{present}} 
\date{January 12, 1999} 
\maketitle

\begin{abstract}
In Landau's phenomenological Fermi-liquid theory (FLT), most physical
quantities are derived from the knowledge of the energy variation $\delta
E\lbrack\delta n\rbrack$ corresponding to a change $\delta n$ of the
quasi-particle (QP) distribution function $n\equiv \lbrace n_{{\bf k}\sigma
}\rbrace$. We show that the internal energy $E\lbrack n\rbrack$ (or,
more precisely, the thermodynamic potential $\Phi\lbrack n\rbrack$),
expressed as a function 
of the QP distribution $n$, can be interpreted as an {\it effective
potential} (in the sense of field theory), which is obtained
from the free energy by a Legendre transformation. 
This allows to obtain explicitly $\delta\Phi$ (or $\delta E$)
starting from a microscopic Hamiltonian and to relate the Landau $f$
function to the forward-scattering two-particle vertex without considering the
collective modes as in the standard diagrammatic derivation of FLT. 
Out-of-equilibrium properties are obtained by extending the definition of
the effective potential  to space- and time-dependent
configurations. $\Phi\lbrack n\rbrack$ is then a functional of the Wigner
distribution function $n\equiv \lbrace n_{{\bf k}\sigma}({\bf
r},t)\rbrace$. It contains information about both the static and
dynamic properties of the Fermi liquid. In particular, it yields the
quantum Boltzmann equation satisfied by $n_{{\bf k}\sigma}({\bf
r},t)$. Finally, we show how $\delta\Phi\lbrack\delta n\rbrack$ can be
derived (in the static case) using a 
finite-temperature renormalization-group approach. In agreement with
previous results based on this technique, we find that the Landau $f$
function is defined by the fixed-point value of the $\Omega$-limit of
the forward-scattering two-particle vertex. 
\end{abstract} 

]

\section{Introduction}

Landau's original approach to Fermi-liquid theory (FLT) is phenomenological.
\cite{Landau57} The main assumption is the existence of a one-to-one 
correspondence between the low-energy elementary excitations 
(quasi-particles (QP's)) of the Fermi liquid and  
the elementary excitations (`particles' and `holes')
of the non-interacting fermion gas. Thermodynamic quantities are
derived from the knowledge of the energy variation $\delta E[\delta
n]$ corresponding to a change $\delta n$ of the QP distribution
function $n=\lbrace n_{\bf k\sigma }\rbrace$. [$\bf k$ and $\sigma $
denote the QP momentum and spin.] $\delta E[\delta n]$ is parameterized by  
the Landau $f$ function (or, equivalently, the Landau parameters) which 
describes the interaction between two QP's. Non-equilibrium properties
are obtained by generalizing the energy functional $\delta E[\delta
n]$ to space- and time-dependent configurations, and by simultaneously
considering the Boltzmann transport equation satisfied by the QP
distribution.  

Much effort has been dedicated to justify Landau's phenomenological 
FLT from a microscopic Hamiltonian. 
The standard derivation consists in showing that the QP dynamics
obtained from the microscopic two-particle vertex agrees with
the conclusion of the phenomenological
FLT. \cite{Landau59,Abrikosov63} Although this
demonstration provides a microscopic definition of the Landau $f$
function, it does not aim at calculating explicitly the energy
functional $\delta E[\delta n]$. 

The so-called statistical FLT provides an alternative approach which
is in spirit much closer to the phenomenological theory of Landau. 
The main idea is to express the thermodynamic potential $\Phi[n]$
obtained from a microscopic Hamiltonian as a function of QP occupation
numbers (see Refs.~\onlinecite{Balian61,Balian64,Luttinger68} and references
therein). The equilibrium QP  
distribution function is determined by minimizing the thermodynamic potential.
The excitation energies are defined as functional derivatives of the 
energy with respect to the distribution function.
They are usually referred to as statistical QP energies to distinguish
them from the QP energies obtained from the poles of the
single-particle Green's function. \cite{Balian64}
Although this formulation provides a natural bridge between microscopic 
models and phenomenological FLT, few explicit calculations have been done
in this framework. For instance, the Landau $f$ function (which can be
obtained as the second functional derivative of the energy with 
respect to the QP distribution function) has not been 
related to the forward-scattering two-particle vertex as was done
within the standard FLT.\cite{Landau59,Abrikosov63} Besides, it is not
clear how the collective modes of the Fermi liquid can be studied
within this approach.  

In this paper we propose a unified description of static and dynamic
properties of Fermi liquids. The main idea is to interpret the
thermodynamic potential $\Phi[n]$, expressed in terms of the QP
distribution function $n$, as an {\it effective potential} (in the
sense of field theory), which is obtained from the free energy by a
Legendre transformation. A similar definition of the thermodynamic
potential $\Phi[n]$ can be found in the statistical FLT and in
particular in the work of Balian and De Dominicis.\cite{Balian61}
This allows to obtain explicitly $\delta\Phi$ (or $\delta E$)
starting from a microscopic Hamiltonian and to relate the Landau $f$
function to the forward-scattering two-particle vertex
without considering the 
collective modes as in the standard diagrammatic derivation of FLT. 
Dynamic properties are obtained by extending the definition of
the effective potential  to space- and time-dependent
configurations. $\Phi[n]$ is then a functional of the Wigner distribution
function $n\equiv \lbrace n_{{\bf k}\sigma}({\bf r},t)\rbrace$. We
show how we can extract from $\Phi$ both the response functions of the
Fermi liquid and the quantum Boltzmann equation satisfied by $n_{{\bf
k}\sigma}({\bf r},t)$. 

\subsection*{Outline of the paper}

In section \ref{sec:pFLT}, we briefly review some basic results of the
phenomenological FLT that are useful in subsequent sections. The
microscopic definition of the thermodynamic potential $\Phi[n]$ is
given in section \ref{sec:md}. We first consider the free energy in the
presence of an external source field $h_{{\bf k}\sigma }$ that couples to
the QP number operator. The QP distribution function $n$ is then
obtained by taking the functional derivative of the free energy with
respect to the source field. By performing a Legendre transformation, 
we obtain a functional $\Phi [n]$ of the QP distribution $n$. In field
theory, $\Phi[n]$ is known as an effective potential. The energy 
variation $\delta E[\delta n]$, or equivalently the Landau $f$ function, 
is obtained from the variation $\delta \Phi [\delta n]$ of the thermodynamic 
potential $\Phi [n]$ around its equilibrium value. As a first example, 
we calculate $\Phi [n]$  exactly for non-interacting 
fermions. In the case of interacting fermions, we use general properties of
Legendre transforms to express $\delta \Phi [\delta n]$ as a function of 
the (linear) response function $\chi_{\sigma \sigma'}({\bf k},{\bf k}')$ to
the external field $h_{{\bf k}\sigma}$. We give a microscopic
definition of the Landau $f$ function in terms of the
susceptibility $\chi$. Using the usual assumptions of FLT, we
calculate $\chi$ and recover the standard microscopic definition of $f$. 

In section \ref{sec:dyna}, we extend the definition of the effective
potential $\Phi[n]$ to space- and time-dependent configurations by
considering the Wigner distribution function $n\equiv \lbrace
n_{{\bf k}\sigma}({\bf r},t)\rbrace$. Restricting ourselves to small
deviations $\delta n_{{\bf k}\sigma}({\bf r},t)$ from
equilibrium, we are able to compute the corresponding variation
$\delta\Phi [\delta n]$ of the effective potential. 
In the static and uniform limit
($n_{{\bf k}\sigma}({\bf r},t)=n_{{\bf k}\sigma}$), $\delta\Phi$
reduces to the Landau's functional studied in section \ref{sec:md}. In
out-of-equilibrium cases, by requiring $\delta\Phi$ to be stationary
with respect to variations of the Wigner distribution function $\delta
n_{{\bf k}\sigma}({\bf r},t)$, we obtain the quantum Boltzmann
equation satisfied by  $\delta n_{{\bf k}\sigma}({\bf
r},t)$. We also show how the charge and spin response functions
are related to the effective potential.

In section \ref{sec:RG}, we use a fi\-nite-tem\-pera\-ture
re\-norma\-liza\-tion-group (RG) approach to reproduce the results of section 
\ref{sec:md}. This approach provides an alternative method for
computing the Landau's functional $\delta\Phi[\delta n]$ and confirms
previous results based 
on this technique. \cite{Chitov95,Dupuis96,Chitov98,Dupuis98} We find that the
$f$ function is given by the fixed-point value of the $\Omega$-limit
of the forward-scattering two-particle vertex ($\Gamma^{\Omega*}$). 

Unless otherwise specified, we consider a three-dimensional
spin-$\frac{1}{2}$ fermion gas (of volume $\nu $) with a spherical Fermi
surface, and assume isotropy in spin space. Only short-range repulsive
interactions are taken into account. We use a grand canonical
formalism at finite temperature $T$, the limit $T\to 0$ being taken at
the end of the calculations. We take $\hbar=k_B=1$ throughout the paper.

\section{Phenomenological FLT}
\label{sec:pFLT}

In this section, we summarize the basic results of the phenomenological 
FLT which are useful for our purpose.

\subsection{Landau's functional $\delta\Phi[\delta n]$}

In the ground state of a Fermi liquid, the QP distribution function 
corresponds to a filled Fermi sea: $n^{(0)}_{{\bf k}\sigma }=\Theta (k_F-k)$,
where $k_F$ is the Fermi momentum and $\Theta (x)$ the step function. In the
phenomenological FLT, one postulates that a change $\delta n_{{\bf k}\sigma }
=n_{{\bf k}\sigma }-n^{(0)}_{{\bf k}\sigma }$ of the QP distribution induces 
the energy variation \cite{Landau57,Landau59,Abrikosov63}
\begin{eqnarray}
\delta E [\delta n]&=& \sum _{{\bf k},\sigma } 
\epsilon _{\bf k}\delta n_{{\bf k}\sigma }\nonumber \\ &&
+{1 \over {2\nu }} \sum _{{\bf k},{\bf k'},\sigma ,\sigma '}
f_{\sigma\sigma '}({\bf k},{\bf k'})
\delta n_{{\bf k}\sigma } \delta n_{{\bf k'}\sigma '}, 
\label{deltaE}
\end{eqnarray}
neglecting terms of order $ O(\delta n^3)$. $\epsilon _{\bf k}$ is the energy
of a QP in the absence of other excited QP's
(for simplicity, we consider only cases where $\epsilon_{\bf k}$ does not
depend on spin). In the vicinity of the Fermi surface, it can be
written as  $\epsilon _{\bf k}\simeq v_F^*(k-k_F)+\mu $, where $v_F^*$
is the QP Fermi velocity and $\mu$ the chemical potential. The term of
order $\delta n ^2$ in (\ref{deltaE}) comes from the interaction
between quasi-particles. For states near the Fermi surface, $k\simeq
k_F$ and $k'\simeq k_F$, the function $f_{\sigma\sigma '}({\bf k},{\bf k}')=
f_{\sigma\sigma '}(\theta)$ depends only on the angle 
$\theta$ between ${\bf k}$ and ${\bf k}'$.
The Landau parameters $F_l^s$ and $F_l^a$ are defined by expanding $f$
on the basis of Legendre polynomials:
\begin{equation}
2N(0)f_{\sigma \sigma'}(\theta)=
\sum _{l=0}^\infty (F_l^s+\sigma\sigma'F_l^a)P_l(\cos\theta),
\label{LPdef}
\end{equation} 
where $N(0)=k_F^2/2\pi^2v_F^*$ is the density of states per spin at the
Fermi level.  

In the grand-canonical ensemble at finite temperature, the physical quantity 
of interest is the thermodynamic potential $\Phi [n]=E[n]-\mu N[n]-\beta ^{-1}
S[n]$. $N[n]=\sum_{{\bf k},\sigma}n_{{\bf k}\sigma}$ is the total QP number, 
$S[n]$ the entropy,
and $\beta=1/T$ the inverse temperature. Because of the correspondence 
between excitations of the non-interacting fermion gas and  QP excitations, 
the entropy has the same expression as in a perfect Fermi gas:
\cite{Landau57,Landau59,Abrikosov63}
\begin{equation}
S[n]=-\sum _{{\bf k},\sigma } [n_{{\bf k}\sigma }\ln n_{{\bf k}\sigma }+
(1-n_{{\bf k}\sigma })\ln (1-n_{{\bf k}\sigma })].
\end{equation}
At equilibrium, the QP distribution function $\bar n$ is obtained from the 
condition $\delta \Phi [n]/\delta n_{{\bf k}\sigma }|_{\bar n}=0$:
\begin{equation}
\bar n_{{\bf k}\sigma}=n_F(\tilde \epsilon_{\bf k} -\mu),
\end{equation}
where $n_F(x)=(e^{\beta x}+1)^{-1}$ is the Fermi-Dirac factor, and
\begin{eqnarray}
\tilde \epsilon_{\bf k}&=& \frac{\delta E[n]}{\delta n_{{\bf
k}\sigma}}\Biggl |_{\bar n} \nonumber \\ &=&\epsilon_{\bf k}+\frac{1}{\nu}
\sum _{{\bf k}',\sigma'} f_{\sigma\sigma'}({\bf k},{\bf k}')[\bar n_{{\bf k}'
\sigma'}-n^{(0)}_{{\bf k}'\sigma'}]
\end{eqnarray}
is the QP energy corresponding to the equilibrium distribution
function $\bar n$. Thus, if we expand $\Phi [\bar n+\delta n]=\Phi [\bar n]+
\delta \Phi [\delta n]$ around its equilibrium value, we obtain to
lowest order in $\delta n$ 
\begin{eqnarray}
\delta \Phi [\delta n]&=& 
\frac{1}{2}\sum _{{\bf k},{\bf k}',\sigma,\sigma'}
\Bigl [-
\frac{\delta _{\sigma,\sigma'}\delta _{{\bf k},{\bf k}'}}{n_F'(\tilde 
\epsilon _{\bf k}-\mu)} \nonumber \\ && 
+\frac{1}{\nu} f_{\sigma\sigma'}({\bf k},{\bf k}') \Bigr ] 
\delta n_{{\bf k}\sigma}\delta n_{{\bf k}'\sigma'} ,
\label{dPhi}
\end{eqnarray}
where $n_F'(x)=-\beta/4\cosh^2(\beta x/2)$. 
There is no linear term in $\delta n$ since $\Phi[n]$ is stationary at
equilibrium.

\subsection{Deformation of the Fermi surface}
\label{subsec:dffs}

Because of the
thermal factor $1/n_F'(\epsilon_{\bf k}-\mu)$ in (\ref{dPhi}), 
small variations of the thermodynamic potential correspond to QP excitations 
lying in the thermal broadening of the Fermi surface ($|\tilde\epsilon_{\bf k}
-\mu|\lesssim T$). When $T\to 0$, these excitations have vanishing 
energies and can be viewed as resulting from a displacement $u_\sigma
({\bf \hat k})$ of the Fermi surface. Here ${\bf \hat k}={\bf k}/k$ is
a unit vector in the direction of $\bf k$. It is sometimes convenient
to express $\delta \Phi$ directly in terms of $u_\sigma({\bf \hat
k})$. Writing \cite{note7}
\begin{equation}
\delta n_{{\bf k}\sigma } = v_F^* u_\sigma({\bf \hat k})
n_F'(\tilde \epsilon_{\bf k}-\mu), 
\end{equation}
we obtain
\begin{eqnarray}
\delta \Phi [u] &=& \frac{\nu {v_F^*}^2N(0)}{2}
\sum _{\sigma,\sigma'} \Bigl \lbrace \delta_{\sigma,\sigma'}
\int \frac{d\Omega_{\bf \hat k}}{4\pi} 
u^2_\sigma({\bf \hat k}) \nonumber \\ && 
+N(0) \int \frac{d\Omega_{\bf \hat k}}{4\pi}
\frac{d\Omega_{{\bf \hat k}'}}{4\pi}
f_{\sigma\sigma'}({\bf \hat k},{\bf \hat k}')
u_\sigma({\bf \hat k})u_{\sigma'}({\bf \hat k}') \Bigr\rbrace
\nonumber  \\ &&
\label{deltaF}
\end{eqnarray}
in the limit $T\to 0$. We have used $\tilde \epsilon_{\bf k}\to 
\epsilon_{\bf k}$ and $n_F'(x)\to -\delta (x)$ when $T\to 0$. 
$\Omega_{\bf \hat k}$ denotes the solid angle in the direction of $\bf
\hat k$. Eq.\ (\ref{deltaF}) was first derived by Pomeranchuk
considering the change in energy $\delta E$ resulting from a Fermi
surface displacement $u_\sigma({\bf \hat k})$ at $T=0$.\cite{Pomeranchuk59}
By requiring $\Phi[u]$ to be minimum for $u=0$, we easily deduce from
(\ref{deltaF}) the stability conditions for a 3D Fermi liquid:
$F_l^s>-2l-1$ and $F_l^a>-2l-1$. \cite{Pomeranchuk59}

\subsection{Dynamic properties}
\label{subsec:dp}

The Landau's functional $\delta\Phi[\delta n]$ allows to compute the
thermodynamic properties of the Fermi liquid, 
but does not contain any information about the QP dynamics. To
obtain the latter, one has to extend the definition of $\delta E$ to
out-of-equilibrium configurations:\cite{Landau59,Abrikosov63}
\begin{eqnarray}
\delta E[\delta n^{\rm cl}]&=& \sum _{{\bf k},\sigma } \int d^3r 
\epsilon _{\bf k}\delta n^{\rm cl}_{{\bf k}\sigma }({\bf r},t)
\nonumber \\ &&
+{1 \over {2\nu }} \sum _{{\bf k},{\bf k'},\sigma ,\sigma '}
\int d^3r f_{\sigma\sigma '}({\bf k},{\bf k'})
\nonumber \\ && \times 
\delta n^{\rm cl}_{{\bf k}\sigma }({\bf r},t) 
\delta n^{\rm cl}_{{\bf k'}\sigma '}({\bf r},t).
\label{deltaE1}
\end{eqnarray}
$n^{\rm cl}_{{\bf k}\sigma }({\bf r},t)=n_{{\bf k}\sigma}^{(0)}+
\delta n^{\rm cl}_{{\bf k}\sigma }({\bf r},t)$ gives the probability
of finding a QP with momentum $\bf k$ and spin $\sigma$ at point $\bf
r$ in space and time $t$. We use the notation $n^{\rm cl}$ to
emphasize that this approach is semiclassical, since it is assumed
that one can simultaneously specify the momentum and position of the
QP. For low-energy excitations, the interaction between QP's (last
term of Eq.~(\ref{deltaE1})) can be assumed to be local in space.
The time-dependent QP energy is defined by 
\begin{eqnarray}
\epsilon_{\bf k}({\bf r},t)&=& \frac{\delta E[n^{\rm cl}]}{\delta
n^{\rm cl}_{{\bf 
k}\sigma}({\bf r},t)}\Biggl |_{n^{\rm cl}} \nonumber \\
&=& \epsilon_{\bf k}+\frac{1}{\nu}
\sum _{{\bf k}',\sigma'} f_{\sigma\sigma'}({\bf k},{\bf k}')\delta
n^{\rm cl}_{{\bf k}'\sigma'}({\bf r},t).
\end{eqnarray}

Eq.~(\ref{deltaE1}) is supplemented with the Boltzmann transport equation
\cite{Landau59,Abrikosov63} 
\begin{eqnarray}
\frac{\partial n^{\rm cl}_{{\bf k}\sigma}({\bf r},t)}{\partial t} 
-\mbox{\boldmath$\nabla$}_{\bf k}n^{\rm cl}_{{\bf k}\sigma}({\bf r},t)
\cdot \mbox{\boldmath$\nabla$}_{\bf r}\epsilon_{\bf k}({\bf r},t)
&& \nonumber \\ 
+\mbox{\boldmath$\nabla$}_{\bf r}n^{\rm cl}_{{\bf k}\sigma}({\bf r},t)
\cdot \mbox{\boldmath$\nabla$}_{\bf k}\epsilon_{\bf k}({\bf r},t)
&=&0,
\label{tr1}
\end{eqnarray}
where we have used the quasi-classical equations $d{\bf r}/dt=
\mbox{\boldmath$\nabla$}_{\bf k}\epsilon_{\bf k}({\bf r},t)$
and $d{\bf k}/dt= -\mbox{\boldmath$\nabla$}_{\bf
r}\epsilon_{\bf k}({\bf r},t)$. To first order in $\delta
n^{\rm cl}$, Eq.~(\ref{tr1}) reduces to 
\begin{eqnarray}
\frac{\partial \delta n^{\rm cl}_{{\bf k}\sigma}({\bf r},t)}{\partial t}
+ v_F^*{\bf\hat k}\cdot \mbox{\boldmath$\nabla$}_{\bf r}\delta n^{\rm
cl}_{{\bf k}\sigma}({\bf r},t) \nonumber && \\ 
+ \delta (\epsilon_{\bf k}-\mu)\frac{1}{\nu}\sum_{{\bf k}',\sigma'}
f_{\sigma\sigma'}({\bf k},{\bf k}') v_F^*{\bf\hat k} \cdot 
\mbox{\boldmath$\nabla$}_{\bf r}\delta n^{\rm cl}_{{\bf k}'\sigma'}({\bf r},t)
&=& 0. \nonumber  \\ &&
\label{tr2}
\end{eqnarray}
The solution of this equation can be written as 
\begin{equation}
\delta n^{\rm cl}_{{\bf k}\sigma}({\bf r},t) = v_F^* u_\sigma({\bf\hat
k},q,\Omega)\delta (\epsilon_{\bf k}-\mu)e^{i({\bf q}\cdot {\bf r}-\Omega t)} 
\end{equation}
where $u_\sigma({\bf\hat k},q,\Omega)$ is a dynamic displacement of the Fermi
surface. The solutions of  Eq.~(\ref{tr2}) correspond to zero-sound
($u_\uparrow=u_\downarrow$) and spin-wave ($u_\uparrow=-u_\downarrow$)
modes.

\section{Landau's functional $\Phi $ as an effective potential}
\label{sec:md}

In statistical mechanics and in the theory of phase transitions, it is
natural to consider the Legendre transform of the free energy. In
magnetism for instance, the Gibbs thermodynamic potential, which is a
functional of the magnetization ${\bf M}(\bf r)$, is obtained from the
free energy by performing a Legendre transformation. \cite{Lebellac}
In field theory, such Legendre transforms are known as effective
potentials.\cite{Negele}
In this section we show that the Landau's functional $\Phi[n]$ can be
naturally interpreted as an effective potential. 

\subsection{Effective potential $\Phi$}

In order to define the effective potential $\Phi$, we first calculate
the free energy $-\beta^{-1} \ln Z[h]$ in presence of a (real) external source
field $h_{{\bf k}\sigma}$ that couples to the QP occupation
number operator $\hat n_{{\bf k}\sigma}$. Then we obtain the Landau's
functional $\Phi[n]$ by performing a Legendre transformation. 

We write the partition function in the Matsubara formalism as a
functional integral over Grassmann variables $\psi ^{(*)} _\sigma
({\bf k},\tau)$ ($\tau $ is an imaginary time):
\begin{equation}
Z[h]=\int {\cal D}\psi ^*{\cal D}\psi \, e^{-S[\psi^*,\psi]-S_h[\psi^*,\psi]},
\label{Z1}
\end{equation}
where $S[\psi^*,\psi]$ is the action when $h_{{\bf k}\sigma }=0$. The
source field contributes to the action a term 
\begin{equation}
S_h[\psi^*,\psi]=\sum _{{\bf k},\sigma }h_{{\bf k}\sigma }
\int _0^\beta d\tau \, \hat n_{{\bf k}\sigma }(\tau ).
\label{Sh0}
\end{equation}
Note that we do not specify at this point
the expression of $\hat n_{{\bf k}\sigma }(\tau)$ as a function of the $\psi$
field. Only for non-interacting fermions do we have the simple relation 
$\hat n_{{\bf k}\sigma }(\tau)=\psi ^*_\sigma ({\bf k},\tau)\psi 
_\sigma ({\bf k},\tau)$. The QP occupation number 
$n_{{\bf k}\sigma }=\langle \hat n_{{\bf k}\sigma }\rangle $ is
obtained from the functional derivative of the free energy:
\begin{equation}
n_{{\bf k}\sigma }=-\frac{1}{\beta } \frac{\delta \ln Z[h]}{\delta h_{{\bf k}
\sigma }}.
\label{ndef}
\end{equation}

Now we introduce the Legendre transform \cite{BdD}
\begin{equation}
\Phi [n]=-\frac{1}{\beta }\ln Z[h] -\sum _{{\bf k},\sigma }h_{{\bf k}\sigma }
n_{{\bf k}\sigma } ,
\label{Phi1}
\end{equation}
where $h_{{\bf k}\sigma }[n]$ is obtained by inverting (\ref{ndef}). The 
equation of state for the thermodynamic potential $\Phi [n]$ reads 
\begin{equation}
 \frac{\delta \Phi[n]}{\delta n_{{\bf k}\sigma }}=-h_{{\bf k}\sigma }.
\label{hdef}
\end{equation}
At equilibrium, i.e., in the absence of source field, $\Phi [n]$ is stationary
with respect to small variations of the QP distribution function. 

\subsection{Non-interacting fermions}

As a first example, we consider non-interacting fermions.
Using $\hat n_{{\bf k}\sigma }(\tau)=\psi ^*_\sigma ({\bf k},\tau )
\psi _\sigma ({\bf k},\tau )$, we obtain: 
\begin{eqnarray}
Z[h] &=& \int {\cal D}\psi ^*{\cal D}\psi \, e^{\sum _{{\bf k},
\sigma ,\omega } \psi ^*_\sigma ({\bf k},\omega )(i\omega -\epsilon _{\bf k}-
h_{{\bf k}\sigma }+\mu )\psi _\sigma ({\bf k},\omega )} \nonumber \\
&=& \prod _{{\bf k},\sigma }\Bigl (1+e^{-\beta (\epsilon _{\bf k} 
+h_{{\bf k}\sigma }-\mu )} \Bigr ),
\label{Z0}
\end{eqnarray}
where $\epsilon_{\bf k}$ is the energy of a fermion with momentum $\bf
k$. We have introduced the Fourier transformed field $\psi ^{(*)}_\sigma 
({\bf k},\omega)$ where $\omega =\pi T(2m+1)$ ($m$ integer) is a fermionic 
Matsubara frequency. Eq.~(\ref{ndef}) yields 
\begin{equation}
n_{{\bf k}\sigma }=n_F(\epsilon _{\bf k} +h_{{\bf k}\sigma }-\mu ).
\label{nff}
\end{equation}
Inverting (\ref{nff}) and using (\ref{Z0}), we eventually obtain
\begin{eqnarray}
\Phi [n] &=& \sum _{{\bf k},\sigma }(\epsilon _{\bf k}-\mu )n_{{\bf k}\sigma }
+\frac{1}{\beta } \sum _{{\bf k},\sigma } [n_{{\bf k}\sigma }
\ln n_{{\bf k}\sigma } \nonumber \\ && + (1-n_{{\bf k}\sigma })
\ln (1-n_{{\bf k}\sigma })].
\label{Pnif}
\end{eqnarray}
Eq.~(\ref{Pnif}) is the expected result for non-interacting fermions.

\subsection{Interacting fermions}

For interacting fermions, it is not possible to calculate exactly the
thermodynamic potential $\Phi [n]$. However, we do not require the whole 
knowledge of $\Phi [n]$, but only its variation $\delta 
\Phi [\delta n]$ when the QP distribution function $n$ varies from its
equilibrium value $\bar n=n|_{h=0}$ by an amount $\delta n$. [For $T\to 0$, 
$\bar n_{{\bf k}\sigma }=\Theta (k_F-k)$ corresponds to the ground-state
distribution function.] This turns out to be a much easier task than obtaining
$\Phi[n]$. 

Expanding $\Phi [\bar n+\delta n]$ to second order in $\delta n$, we obtain
\begin{equation}
\delta \Phi [\delta n]=\frac{1}{2}
\sum _{{\bf k},{\bf k}',\sigma,\sigma '}
\frac{\delta ^{(2)}\Phi [n]}{\delta n_{{\bf k}\sigma }
\delta n_{{\bf k}'\sigma '}} \Biggl | _{\bar n} 
\delta n_{{\bf k}\sigma }\delta n_{{\bf k}'\sigma '}.
\end{equation}
There is no linear term since $\Phi[n]$ is stationary at equilibrium 
[Eq.~(\ref{hdef})]. To proceed further, we use the following relation
between the functional derivatives of the free energy $-\beta ^{-1}\ln
Z[h]$ and those of the Legendre transform $\Phi [n]$: 
\cite{Negele}
\begin{equation}
\sum _{{\bf k}_3,\sigma _3} 
\frac{\beta ^{-1}\delta ^{(2)}\ln Z[h]}{\delta h_{{\bf k}_1\sigma_1}
\delta h_{{\bf k}_3\sigma_3}}\Biggl | _{h=0}
\frac{\delta ^{(2)}\Phi [n]}{\delta n_{{\bf k}_3\sigma_3}
\delta n_{{\bf k}_2\sigma_2}} \Biggl | _{\bar n} = \delta _{\sigma_1,\sigma_2}
\delta _{{\bf k}_1,{\bf k}_2}.
\label{chidef}
\end{equation}
Introducing the matrix 
\begin{equation}
\chi _{\sigma\sigma'}({\bf k},{\bf k}')=
\frac{\beta ^{-1}\delta ^{(2)} \ln Z[h]}{\delta h_{{\bf k}\sigma}
\delta h_{{\bf k}'\sigma'}}\Biggl |_{h=0},
\label{chidef1}
\end{equation} 
we thus obtain 
\begin{equation}
\delta \Phi [\delta n]=\frac{1}{2}
\sum _{{\bf k},{\bf k}',\sigma,\sigma '}
\chi _{\sigma\sigma'}^{-1}({\bf k},{\bf k}')
\delta n_{{\bf k}\sigma }\delta n_{{\bf k}'\sigma '}.
\label{dPhi1}
\end{equation}
Note that $\chi_{\sigma\sigma'}({\bf k},{\bf k}')$ is nothing but the
(linear) response function to the external source field $h_{{\bf
k}\sigma }$. Comparing (\ref{dPhi}) and (\ref{dPhi1}) (and assuming
that the system is a Fermi liquid), we obtain a
{\it microscopic definition of the Landau $f$ function}:
\begin{equation}
\frac{1}{\nu} f_{\sigma\sigma'}({\bf k},{\bf k}')=
\frac{\delta _{\sigma,\sigma'}\delta _{{\bf k},{\bf
k}'}}{n_F'(\epsilon_{\bf k}-\mu)}+
\chi_{\sigma\sigma'}^{-1}({\bf k},{\bf k}').
\label{chidef2}
\end{equation}
From now on, we ignore the difference between $\tilde \epsilon_{\bf
k}$ and $\epsilon_{\bf k}$ since we are ultimately interested in the
limit $T\to 0$. Thus, the calculation of the Landau $f$ function reduces to the
calculation of the susceptibility $\chi $. In the next section, we
show how $\chi$ can be calculated using the standard assumptions of
FLT.

\subsection{Calculation of the Landau $f$ function}
\label{subsec:func}

In this section, we first express the Landau $f$ function in terms of the
QP properties, and then relate it to the bare fermion properties.

\subsubsection{$f$ in terms of QP properties}

Since the external field $h_{{\bf k}\sigma}$ couples to the QP's, we
have to distinguish between the QP field (that we denote by $\bar
\psi^{(*)}$) and the field $\psi^{(*)}$ corresponding to the bare
fermions. With these notations, we have $\hat n_{{\bf k}\sigma}(\tau)
=\bar\psi^*_\sigma({\bf k},\tau)\bar\psi_\sigma({\bf k},\tau)$, where
the operator $\hat n_{{\bf k}\sigma}(\tau)$ has been introduced in
(\ref{Sh0}). From (\ref{Sh0}) and (\ref{chidef1}), we obtain 
\begin{equation}
\chi_{\sigma\sigma'}({\bf k},{\bf k}';\tilde q)=\frac{1}{\beta}\sum
_{\omega,\omega'} \tilde \chi _{\sigma\sigma'}(\tilde k,\tilde
k';\tilde q),
\label{chitilde}
\end{equation}
where $\tilde \chi$ is the two-particle Green's function 
\begin{eqnarray}    
\tilde \chi_{\sigma\sigma'}(\tilde k,\tilde k';\tilde q)
&=& \langle \bar \psi ^*_\sigma (\tilde k_-)\bar \psi_\sigma
(\tilde k_+) \bar \psi ^*_{\sigma'}(\tilde k'_+)\bar
\psi_{\sigma'}(\tilde k'_-) \rangle_c \nonumber \\ 
&\equiv& \langle \bar \psi ^*_\sigma (\tilde k_-)\bar \psi_\sigma
(\tilde k_+) \bar \psi ^*_{\sigma'}(\tilde k'_+)\bar
\psi_{\sigma'}(\tilde k'_-) \rangle  
\nonumber \\ && - \delta_{\tilde q,0} 
\langle\bar\psi^*_\sigma(\tilde k)\bar\psi_\sigma (\tilde k)\rangle 
\langle\bar\psi^*_{\sigma'}(\tilde k')\bar\psi_{\sigma'}(\tilde
k')\rangle . \nonumber \\ && 
\label{chitilde1}
\end{eqnarray}
The average values are evaluated in the absence of source field ($h=0$),
and the meaning of $\langle\cdots\rangle_c$ is defined by the second
line of (\ref{chitilde1}). We use the notation $\tilde k=({\bf
k},i\omega)$, $\tilde q=({\bf q},i\Omega)$, and $\tilde
k_{\pm}=\tilde k\pm \tilde q/2$. $\omega$ and $\Omega$ denote
fermionic and bosonic Matsubara frequencies, respectively. 
The Landau $f$ function is related to the limit 
$\tilde q\to 0$ of $\chi_{\sigma\sigma'}({\bf k},{\bf k}';\tilde
q)$ [Eq.~(\ref{chidef2})].
In a Fermi liquid, the forward-scattering limit $\tilde q\to 0$
is ill-defined and one should distinguish between the $q$- and
$\Omega$-limits, which correspond to $\Omega/q\to 0$ and $q/\Omega \to 0$,
respectively. Since a static and uniform external field cannot create
quasi-particle-quasi-hole excitations (and therefore modify the
ground-state QP distribution function) in the $\Omega$-limit,\cite{note4}
$f$ is obtained from the $q$-limit of the function 
$\chi$. It turns out to be more convenient to keep a finite
$\tilde q$ at intermediate stages, the limit $\tilde q\to 0$ (with
$\Omega/q\to 0$) being taken only at the end of the
calculations. Besides, the function $\chi$ (with $\tilde q$ finite)
will also have to be considered in the analysis of the dynamic
properties of Fermi liquids (section \ref{sec:dyna}). 

$\tilde \chi$ satisfies the Dyson equation
\begin{eqnarray}
\tilde \chi_{\sigma\sigma'}(\tilde k,\tilde k';\tilde q)&=&
\tilde \chi^{(0)}_{\sigma\sigma'}(\tilde k,\tilde k';\tilde q)
\nonumber \\ &&
-\frac{1}{\beta \nu}\sum_{\tilde k_1,\tilde k_2,\sigma_1,\sigma_2}
\tilde \chi^{(0)}_{\sigma\sigma_1} (\tilde 
k,\tilde k_1;\tilde q) \nonumber \\ && \times 
\bar \Gamma _{\sigma_1\sigma_2}^{\rm irr}(\tilde k_1,\tilde k_2;\tilde
q) \tilde \chi_{\sigma_2\sigma'}(\tilde k_2,\tilde k';\tilde q),
\label{Dyson}
\end{eqnarray}
where 
\begin{equation}
\bar \Gamma^{\rm irr}_{\sigma\sigma'}(\tilde k,\tilde k';\tilde q)\equiv
\bar \Gamma^{\rm irr}_{\sigma\sigma',\sigma'\sigma}
(\tilde k_+,\tilde k'_-,\tilde k'_+,\tilde k_-) 
\end{equation}
is the irreducible QP vertex in the Landau channel. \cite{note6} The
non-interacting part 
\begin{equation}
\tilde \chi ^{(0)}_{\sigma\sigma'}(\tilde k,\tilde k';\tilde q)=-\delta
_{\sigma,\sigma'}\delta _{\tilde k,\tilde k'} 
\bar G(\tilde k_+)\bar G(\tilde k_-)
\end{equation}
is easily expressed in terms of the QP propagators
\begin{eqnarray}
\bar G(\tilde k_\pm)&=&-\langle \bar \psi_\sigma(\tilde
k_\pm)\bar \psi_\sigma^*(\tilde k_\pm)\rangle \nonumber \\ &=&
\Bigl \lbrack i\omega \pm i\frac{\Omega}{2}-\epsilon_{\bf k}\mp {\bf
v}_{\bf k}\cdot \frac{\bf q}{2}+\mu \Bigr \rbrack ^{-1} , 
\end{eqnarray}
where $\epsilon_{\bf k}$ is the QP energy and ${\bf v}_{\bf
k}= \mbox{\boldmath$\nabla$} _{\bf k}\epsilon _{\bf k}$ the QP group velocity. 
Approximating $\epsilon_{\bf k}$ by $v_F^*(k-k_F)+\mu$, we have ${\bf
v}_{\bf k}=v_F^*{\bf \hat k}$ (with ${\bf \hat k}={\bf k}/k$). 
The quantity $\tilde \chi^{(0)}$ becomes singular in the
forward-scattering limit $\tilde q\to 0$ since the poles of the two QP
propagators coalesce. In FLT, one assumes that $\tilde \chi^{(0)}$ is
the only singular quantity in this limit. 
\cite{Landau59,Abrikosov63}
This implies that the irreducible vertex $\bar
\Gamma^{\rm irr}$, which does not contain $\tilde \chi^{(0)}$, is a
non-singular quantity that has a well-defined limit when $\tilde q\to
0$. We therefore set $\tilde q=0$ in $\bar \Gamma ^{\rm irr}$. 
Furthermore, since the singularity of $\tilde \chi^{(0)}$ is due 
to QP states in the vicinity of the Fermi surface,
we can ignore the $k$ and $\omega$ dependence of  $\bar
\Gamma^{\rm irr}$ (i.e., set $k=k'=k_F$ and $\omega=\omega'=0$), which
then becomes a function of ${\bf \hat k}$ and ${\bf \hat k}'$. [The
variables $|k-k_F|,|k'-k_F|$ and $\omega,\omega'$ are irrelevant in
the RG sense (see section \ref{sec:RG}).] This allows to perform the 
frequency sums in (\ref{Dyson}) and obtain a Dyson equation for $\chi$: 
\begin{eqnarray}
\chi_{\sigma\sigma'}({\bf k},{\bf k}';\tilde q)&=&
\chi^{(0)}_{\sigma\sigma'}({\bf k},{\bf k}';\tilde q)
\nonumber \\ &&
-\frac{1}{\nu}\sum_{{\bf k}_1,{\bf k}_2,\sigma_1,\sigma_2} 
\chi^{(0)}_{\sigma\sigma_1}({\bf k},{\bf k}_1;\tilde q)
\nonumber \\ && \times 
\bar \Gamma _{\sigma_1\sigma_2}^{\rm irr}({\bf \hat k}_1,{\bf \hat k}_2) 
\chi_{\sigma_2\sigma'}({\bf k}_2,{\bf k}';\tilde q).
\label{Dyson1}
\end{eqnarray}
This can be rewritten as 
\begin{equation}
\chi^{-1}_{\sigma\sigma'}({\bf k},{\bf k}';\tilde q)=
\chi^{(0)-1}_{\sigma\sigma'}({\bf k},{\bf k}';\tilde q)
+\frac{1}{\nu}\bar \Gamma ^{\rm irr}_{\sigma\sigma'}({\bf \hat k},{\bf
\hat k}').
\label{chi52}
\end{equation}
In Eq.~(\ref{chi52}), $\chi^{-1}(\tilde q)$, $\chi^{(0)-1}(\tilde q)$
and $\bar \Gamma^{\rm irr}$ should be understood as matrices in spin
($\sigma$) and momentum ($\bf k$) space. Comparing (\ref{chidef2}) and
(\ref{chi52}), and noting that
\begin{eqnarray}
\chi^{(0)}_{\sigma\sigma'}({\bf k},{\bf k}';\tilde q)&=&
\delta_{\sigma,\sigma'}\delta_{{\bf k},{\bf k}'}
n_F'(\epsilon_{\bf k}-\mu)
\frac{{\bf v}_{\bf k}\cdot {\bf q}}{i\Omega-{\bf v}_{\bf k}\cdot {\bf q}}
\nonumber \\ &\to & 
-\delta_{\sigma,\sigma'}\delta_{{\bf k},{\bf k}'}
n_F'(\epsilon_{\bf k}-\mu) 
\label{chi0}
\end{eqnarray}
in the $q$-limit, we conclude that 
\begin{equation}
f_{\sigma\sigma'}({\bf k},{\bf k}')=\bar \Gamma^{\rm irr}_{\sigma\sigma'}
({\bf \hat k},{\bf \hat k}').
\label{f1}
\end{equation}

We can proceed one step further by relating $\bar\Gamma^{\rm irr}$ to
the total QP vertex $\bar\Gamma$, using the equation
\begin{eqnarray}
\bar \Gamma_{\sigma\sigma'}(\tilde k,\tilde k';\tilde q)&=&
\bar \Gamma^{\rm irr}_{\sigma\sigma'}({\bf \hat k},{\bf \hat k}')
-\frac{1}{\beta \nu} \sum_{\tilde k_1,\tilde k_2,\sigma_1,\sigma_2} 
\bar \Gamma^{\rm irr}_{\sigma\sigma_1} ({\bf \hat k},{\bf \hat k}_1)
\nonumber \\ && \times \tilde \chi^{(0)}_{\sigma_1\sigma_2}(\tilde
k_1,\tilde k_2;\tilde q) \bar \Gamma_{\sigma_2\sigma'}(\tilde
k_2,\tilde k';\tilde q). 
\label{Gbar}
\end{eqnarray}
Since $\bar\Gamma$, as $\bar\Gamma^{\rm irr}$, is independent of
frequencies, we can perform the sum over $\omega_1$ and $\omega_2$ in
(\ref{Gbar}), which yields 
\begin{eqnarray}
\bar \Gamma_{\sigma\sigma'}({\bf \hat k},{\bf \hat k}';\tilde q)&=&
\bar \Gamma^{\rm irr}_{\sigma\sigma'}({\bf \hat k},{\bf \hat k}')
-\frac{1}{\nu} \sum_{{\bf k}_1,{\bf k}_2,\sigma_1,\sigma_2} 
\bar \Gamma^{\rm irr}_{\sigma\sigma_1} ({\bf \hat k},{\bf \hat k}_1)
\nonumber \\ && \times \chi^{(0)}_{\sigma_1\sigma_2}({\bf k}_1,{\bf
k}_2;\tilde q) \bar \Gamma_{\sigma_2\sigma'}({\bf \hat k}_2,{\bf \hat
k}';\tilde q).
\end{eqnarray}
$\chi^{(0)}$ vanishing in the $\Omega$-limit [Eq.~(\ref{chi0})], we
obtain
\begin{eqnarray}
\bar \Gamma_{\sigma\sigma'}^{\rm irr}({\bf \hat k},{\bf \hat k}')&=& 
\lim_{\Omega \to 0} \Bigl [ \bar \Gamma _{\sigma\sigma'}
({\bf \hat k},{\bf \hat k}';\tilde q) \Bigr |_{q=0}
 \Bigr ] \nonumber \\ &\equiv& \bar \Gamma^\Omega 
_{\sigma\sigma'}({\bf \hat k},{\bf \hat k}').
\label{Gbar1}
\end{eqnarray}
Eqs.~(\ref{f1},\ref{Gbar1}) relate the Landau $f$ function to the
$\Omega$-limit of the forward-scattering QP vertex.

\subsubsection{$f$ in terms of bare fermion properties}
\label{subsubsec:Gbar}

The last step of our derivation is to relate $\bar \Gamma ^{\rm irr}$
to the vertex of the bare fermions. Consider first the single-particle
propagator in a Fermi liquid. It can be written as
\cite{Landau59,Abrikosov63}
\begin{eqnarray}
G(\tilde k) &=& -\langle \psi_\sigma(\tilde k)\psi_\sigma^*(\tilde
k)\rangle \nonumber \\ &=& \frac{z}{i\omega-\epsilon_{\bf k}+\mu} +
G_{\rm inc}(\tilde k), 
\end{eqnarray}
where $ G_{\rm inc}(\tilde k)$ is the incoherent part, and $z$ the QP
renormalization factor. The QP field
$\bar \psi^{(*)}$ can be obtained (at least formally) by filtering out
the incoherent part and rescaling the field in order to ensure that
the total spectral weight is equal to one (see Nozi\`eres' book 
\cite{Abrikosov63} for a detailed discussion). [As
discussed in section \ref{sec:RG}, the RG approach provides an
explicit realization of this procedure.] $G$ and $\bar G$ are then related by
\begin{equation}
G(\tilde k) = z \bar G(\tilde k)+ G_{\rm inc}(\tilde k).
\end{equation}

Consider now the two-particle Green's function
\begin{equation}
G^{\rm II}_{\sigma\sigma'}(\tilde k,\tilde k';\tilde q)= 
\langle \psi_\sigma^*(\tilde k_-)\psi_\sigma(\tilde k_+)
\psi_{\sigma'}^*(\tilde k'_+)\psi_{\sigma'}(\tilde k'_-) \rangle _c.
\end{equation}
$G^{\rm II}$ can be written as 
\begin{eqnarray}
G^{\rm II}_{\sigma\sigma'}(\tilde k,\tilde k';\tilde q)&=&
{G^{{\rm II}(0)}_{\sigma\sigma'}}(\tilde k,\tilde k';\tilde q)
\nonumber \\ && 
-\frac{1}{\beta \nu} \sum_{\tilde k_1,\tilde k_2,\sigma_1,\sigma_2} 
{G^{{\rm II}(0)}_{\sigma\sigma_1}}(\tilde k,\tilde k_1;\tilde q)
\nonumber \\ && \times 
\Gamma_{\sigma _1\sigma_2}(\tilde k_1,\tilde k_2;\tilde q)
{G^{{\rm II}(0)}_{\sigma_2\sigma'}}(\tilde k_2,\tilde k';\tilde q),
\label{GIIeq}
\end{eqnarray}
where 
\begin{equation}
\Gamma_{\sigma \sigma'}(\tilde k,\tilde k';\tilde q)\equiv
\Gamma_{\sigma \sigma',\sigma',\sigma}(\tilde k_+,\tilde
k'_-,\tilde k'_+,\tilde k_-) 
\end{equation}
is the total vertex for the bare fermions, and
\begin{eqnarray}
{G^{{\rm II}(0)}_{\sigma\sigma'}}(\tilde k,\tilde k';\tilde q)&=&
-\delta_{\sigma,\sigma'}\delta_{\tilde k,\tilde k'}G(\tilde k_+)G(\tilde k_-)
\nonumber \\ &=& -\delta_{\sigma,\sigma'}\delta_{\tilde k,\tilde k'} \lbrack
z^2 \bar G(\tilde k_+)\bar G(\tilde k_-)+\varphi (\tilde k) \rbrack .
\nonumber \\ 
\label{GII0}
\end{eqnarray}
We have separated the coherent part of the particle-hole pair
propagation from the incoherent part $\varphi (\tilde k)$. Note that
$\varphi (\tilde k)$ is non-singular in the limit $\tilde q\to 0$ and
is therefore evaluated at $\tilde q=0$. Inserting (\ref{GII0}) into
(\ref{GIIeq}) and retaining only the 
coherent part of the particle-hole propagation, we obtain an equation
for the Green's function $\tilde \chi$ introduced in section
\ref{subsubsec:Gbar}: 
\begin{eqnarray}
\tilde \chi_{\sigma\sigma'}(\tilde k,\tilde k';\tilde q)&=&
{\tilde \chi_{\sigma\sigma'}}^{(0)}(\tilde k,\tilde k';\tilde q)
\nonumber \\ &&
-\frac{1}{\beta \nu} \sum_{\tilde k_1,\tilde k_2,\sigma_1,\sigma_2} 
{\tilde \chi_{\sigma\sigma_1}}^{(0)}(\tilde k,\tilde k_1;\tilde q)
\nonumber \\ && \times 
z^2\Gamma_{\sigma _1\sigma_2}(\tilde k_1,\tilde k_2;\tilde q)
{\tilde \chi_{\sigma_2\sigma'}}^{(0)}(\tilde k_2,\tilde k';\tilde q).
\label{GIIeq1}
\end{eqnarray}
We then deduce from (\ref{Dyson}) and (\ref{GIIeq1})
\begin{equation}
\bar \Gamma _{\sigma\sigma'}(\tilde k,\tilde k';\tilde q)=
z^2\Gamma_{\sigma\sigma'}(\tilde k,\tilde k';\tilde q).
\label{Gbar2}
\end{equation}
Eq.~(\ref{Gbar2}) relates the QP vertex $\bar \Gamma$ to the bare
fermion vertex $\Gamma$. From (\ref{f1},\ref{Gbar1},\ref{Gbar2}) we
obtain the well-known expression of the Landau $f$ function in a Fermi liquid:
\begin{equation}
f_{\sigma\sigma'}({\bf k},{\bf k}')=z^2 \Gamma^\Omega 
_{\sigma\sigma'}({\bf \hat k},{\bf \hat k}'). 
\end{equation}

\section{Dynamic properties}
\label{sec:dyna}

The Landau's functional $\Phi[n]$ does not contain any information about
the dynamic properties of the Fermi liquid. 
In this section, we extend the definition of the effective potential
$\Phi$ to space- and time-dependent configurations by considering the 
Wigner distribution function (WDF) $n\equiv \lbrace
n_{{\bf k}\sigma}({\bf r},t)\rbrace$. This allows to obtain the QP
dynamics and the collective modes of the Fermi liquid without
introducing the semiclassical distribution function $n^{\rm
cl}_{{\bf k}\sigma}({\bf r},t)$ and the corresponding Boltzmann
transport equation as in the phenomenological FLT (see section
\ref{subsec:dp}).

\subsection{Generalized effective potential $\Phi$}
\label{subsec:tdep}

The best quantum analog to the  semiclassical distribution function
$n^{\rm cl}_{{\bf k}\sigma}({\bf r},t)$ is the WDF. 
\cite{Wigner32} The latter is not a true distribution function 
since it is not positive definite. However, as far as its
moments are concerned, it behaves similarly to a distribution
function. \cite{Wigner32,Mahan} For our purpose, we define the WDF in
Matsubara time as  
\begin{eqnarray}
n_{{\bf k}\sigma}({\bf r},\tau)&=&\int d^3r' e^{-i{\bf k}\cdot {\bf r}'}
\nonumber \\ && \times 
\Bigl \langle \bar\psi_\sigma^*\Bigl ({\bf r}-\frac{{\bf
r}'}{2},\tau\Bigr ) \bar\psi_\sigma\Bigl ({\bf r}+\frac{{\bf
r}'}{2},\tau\Bigr ) \Bigr \rangle ,
\end{eqnarray}
and its Fourier transform as
\begin{eqnarray}
n_{{\bf k}\sigma}(\tilde q) &=& \frac{1}{\beta\nu} \int_0^\beta d\tau
\int d^3r\, e^{-i{\bf q}\cdot {\bf r}+i\Omega\tau} n_{{\bf k}\sigma}
({\bf r},\tau) \nonumber \\ &=& 
\frac{1}{\beta} \sum_\omega \langle \bar\psi_\sigma^*(\tilde k_-)
\bar\psi_\sigma(\tilde k_+) \rangle .
\end{eqnarray} 
Note that these definitions involve the QP field $\bar\psi^{(*)}$. 

We are now in a position to define an effective
potential $\Phi[n]$, which is a functional of the WDF $n\equiv \lbrace n_{{\bf
k}\sigma}(\tilde q)\rbrace$. We shall proceed along the same lines as in
section \ref{sec:md}. We consider the system in presence of a source
field $h_{{\bf k}\sigma}(\tilde q)=h^*_{{\bf k}\sigma}(-\tilde q)$
that couples to the QP operator
$\hat n_{{\bf k}\sigma}(\tilde q)=\beta^{-1}\sum_\omega
\bar\psi_\sigma^*(\tilde k_-) \bar\psi_\sigma(\tilde k_+)$. We write
the partition function as in (\ref{Z1}), with 
\begin{equation}
S_h=\beta \sum_{{\bf k},\sigma,\tilde q} h_{{\bf k}\sigma}(-\tilde q)
\hat n_{{\bf k}\sigma}(\tilde q).
\end{equation}
The WDF is obtained by taking the  functional derivative of the free
energy with respect to the source field:
\begin{equation}
n_{{\bf k}\sigma}(\tilde q)=\langle \hat n_{{\bf k}\sigma}(\tilde
q)\rangle = -\frac{1}{\beta} \frac{\delta \ln Z[h]}{\delta h_{{\bf
k}\sigma} (-\tilde q)}. 
\end{equation}
The effective potential is then defined as 
\begin{equation}
\Phi[n]=-\frac{1}{\beta}\ln Z[h] -\sum_{{\bf k},\sigma,\tilde q}
h_{{\bf k}\sigma}(-\tilde q)n_{{\bf k}\sigma}(\tilde q), 
\end{equation}
and satisfies the `equation of state'
\begin{equation}
\frac{\delta \Phi[n]}{\delta n_{{\bf k}\sigma}(\tilde q)} =- h_{{\bf
k}\sigma}(-\tilde q). 
\label{motion}
\end{equation}
We will show below that in the absence of source field ($h=0$),
the stationarity condition of the effective potential
[Eq.~(\ref{motion})] yields the equation of motion for the WDF. 

Even for non-interacting fermions, $\Phi[n]$ cannot be calculated
exactly. We shall therefore consider only small fluctuations $\delta
n_{{\bf k}\sigma}(\tilde q)$ around the equilibrium state: 
\begin{equation}
n_{{\bf k}\sigma}(\tilde q)=\delta_{\tilde q,0}\bar n_{{\bf k}\sigma} 
+\delta n_{{\bf k}\sigma}(\tilde q).
\end{equation}
Eq.~(\ref{chidef}) can be easily generalized into 
\begin{equation}
\frac{\delta ^{(2)}\Phi[n]}{\delta n_{{\bf k}\sigma}(-\tilde q)
\delta n_{{\bf k}'\sigma'}(\tilde q')} \Biggl |_{\bar n}
= \delta_{\tilde q,\tilde q'}
\chi^{-1}_{\sigma\sigma'}({\bf k},{\bf k}';\tilde q) ,
\label{chi48}
\end{equation}
where 
\begin{equation}
\chi_{\sigma\sigma'}({\bf k},{\bf k}';\tilde q)=
\frac{\beta^{-1}\delta^{(2)}\ln Z[h]}{\delta h_{{\bf k}\sigma}(-\tilde q)
\delta h_{{\bf k}'\sigma'}(\tilde q)}\Biggl |_{h=0}
\end{equation}
is the susceptibility introduced in section \ref{sec:md}
[Eqs.~(\ref{chitilde},\ref{chitilde1})]. The Kronecker symbol
$\delta_{\tilde q,\tilde q'}$ in (\ref{chi48}) results from translational
invariance. To second order in $\delta n$, we thus have
\begin{equation}
\delta\Phi[\delta n]=\frac{1}{2} \sum_{{\bf k},{\bf
k}',\sigma,\sigma',\tilde q} \chi^{-1}_{\sigma\sigma'}({\bf k},{\bf
k}';\tilde q) 
\delta n_{{\bf k}\sigma}(-\tilde q)
\delta n_{{\bf k}'\sigma'}(\tilde q).
\label{Phi30}
\end{equation}
Using Eqs.~(\ref{chi52},\ref{chi0},\ref{f1}), we can write the effective
potential in terms of the Landau $f$ function:
\begin{eqnarray}
\delta\Phi[\delta n]&=&\frac{1}{2} \sum_{{\bf k},{\bf
k}',\sigma,\sigma',\tilde q} \Bigl \lbrace 
\frac{\delta_{\sigma,\sigma'}\delta_{{\bf k},{\bf
k}'}}{n_F'(\epsilon_{\bf k}-\mu)} \frac{i\Omega-{\bf v}_{\bf k}\cdot
{\bf q}}{{\bf v}_{\bf k}\cdot {\bf q}} \nonumber \\ && 
+ \frac{1}{\nu} f_{\sigma\sigma'}({\bf k},{\bf k}') \Bigr \rbrace 
\delta n_{{\bf k}\sigma}(-\tilde q)
\delta n_{{\bf k}'\sigma'}(\tilde q).
\label{Phi3}
\end{eqnarray}
Eq.~(\ref{Phi3}) generalizes the Landau's functional to space- and
time-dependent 
configurations. As in the static and uniform case studied in section
\ref{sec:md}, $\delta\Phi[\delta n]$ is essentially parameterized by
the Landau $f$
function. We shall show in the next section that $\delta\Phi$
contains all the information about the QP dynamics. Moreover, if we
consider the $q$-limit, the WDF $\lbrace n_{{\bf k}\sigma}({\bf r},t)\rbrace$
reduces to the QP distribution function $\lbrace n_{{\bf
k}\sigma}\rbrace$, and $\Phi[n]$ to the thermodynamic potential
introduced in section \ref{sec:md}. Thus the
effective potential $\Phi$ describes static and dynamic properties
of the Fermi liquid in a unified framework.
This result should be contrasted with the phenomenological FLT
which requires slightly different approaches to deal with static
and dynamic properties (see section \ref{sec:pFLT}), and strongly
relies on semiclassical arguments for the latter.

\subsection{Quantum Boltzmann equation}

In the absence of source field ($h=0$), the stationarity condition
of the effective potential [Eq.~(\ref{motion})] yields 
\begin{eqnarray}
({\bf v}_{\bf k}\cdot {\bf q}-i\Omega) \delta n_{{\bf k}\sigma}(\tilde q)
-{\bf v}_{\bf k}\cdot {\bf q}n_F'(\epsilon_{\bf k}-\mu)&& \nonumber \\ 
\times \frac{1}{\nu} \sum_{{\bf k}',\sigma'} f_{\sigma\sigma'} ({\bf
k},{\bf k}') \delta n_{{\bf k}'\sigma'}(\tilde q)&=&0. 
\label{qbe}
\end{eqnarray}
Eq.~(\ref{qbe}) is the quantum Boltzmann equation for the WDF. [Real-time
quantities are obtained by the usual analytic continuation $i\Omega
\to \Omega+i0^+$.] Note
that in the simple case we are considering here (no external electric
field, small fluctuations around the equilibrium state, etc), it
is identical to the Boltzmann equation (\ref{tr2}) satisfied by the
semiclassical distribution function $n^{\rm cl}_{{\bf k}\sigma}({\bf
r},t)$. \cite{Mahan,Negele1} 

The solution of (\ref{qbe}) can be written as 
\begin{equation}
\delta n_{{\bf k}\sigma}(\tilde q)=v_F^*u_\sigma({\bf \hat k},\tilde
q) n_F'(\epsilon_{\bf k}-\mu) ,
\end{equation}
where $u_\sigma({\bf \hat k},\tilde q)$ is naturally interpreted as
the dynamic Fermi surface displacement for spin $\sigma$ fermions. It
satisfies the equation
\begin{eqnarray}
  ({\bf v}_{\bf k}\cdot {\bf q}-i\Omega)u_\sigma({\bf \hat k},\tilde q) 
&& \nonumber \\
+{\bf v}_{\bf k}\cdot {\bf q}N(0)\sum_{\sigma'} \int
\frac{d\Omega_{{\bf \hat k}'}}{4\pi} f_{\sigma\sigma'}({\bf \hat k},{\bf 
\hat k}')u_{\sigma'}({\bf \hat k}',\tilde q) &=&0 .
\label{motion1}
\end{eqnarray}
Here we have used $n_F'(\epsilon_{\bf k}-\mu)=-\delta(\epsilon_{\bf
k}-\mu)$ when $T\to 0$. Eq.~(\ref{motion1}) yields the zero-sound (for
$u_\uparrow=u_\downarrow$) and spin-wave (for $u_\uparrow=-u_\downarrow$)
modes of the Fermi liquid.\cite{Landau59,Abrikosov63}

It is also possible to express the effective potential directly in
terms of the dynamic Fermi surface displacements:
\begin{eqnarray}
\delta\Phi[u] &=& \frac{\nu{v_F^*}^2N(0)}{2}
\sum_{\sigma,\sigma',\tilde q} 
\Bigl \lbrace \delta_{\sigma,\sigma'} \int \frac{d\Omega_{\bf \hat
k}}{4\pi} \frac{{\bf v}_{\bf k}\cdot {\bf q}-i\Omega}{{\bf v}_{\bf k}\cdot
{\bf q}} \nonumber \\ && \times |u_\sigma({\bf\hat k},\tilde q)|^2 
+ N(0) \int \frac{d\Omega_{\bf \hat k}}{4\pi}
\frac{d\Omega_{{\bf \hat k}'}}{4\pi} f_{\sigma\sigma'}({\bf \hat k},{\bf \hat
k}') \nonumber \\ && \times 
u_\sigma({\bf \hat k},-\tilde q) u_\sigma'({\bf \hat k}',\tilde q)
\Bigr \rbrace .
\label{Phi4}
\end{eqnarray}
Here we have used $u_\sigma({\bf \hat k},-\tilde q)=u^*_\sigma({\bf
\hat k},\tilde q)$. 
The equations of motion (\ref{motion1}) are then directly obtained
from the stationarity condition $\delta\Phi/\delta u_\sigma({\bf\hat
k},-\tilde q)=0$. Note also that (\ref{Phi4}) reduces to (\ref{deltaF})
in the $q$-limit.

\subsection{Response functions} 

Since the effective potential is essentially determined by the
susceptibility $\chi_{\sigma\sigma'}({\bf k},{\bf k}';\tilde q)$, it also
yields the response functions of the Fermi liquid. For instance, the
charge and spin response functions are given by
\begin{eqnarray}
\chi_{\rm ch}(\tilde q)&=&\frac{1}{\nu}\sum_{{\bf k},{\bf k}',\sigma,
\sigma'}\chi_{\sigma\sigma'}({\bf k},{\bf k}';\tilde q), \nonumber \\
\chi_{\rm sp}(\tilde q)&=&\frac{1}{\nu}\sum_{{\bf k},{\bf k}',\sigma,
\sigma'}\sigma\sigma'\chi_{\sigma\sigma'}({\bf k},{\bf k}';\tilde q),
\end{eqnarray}
where $\chi$ is related to the effective potential by Eq.~(\ref{Phi30}). 
Note that when calculating response functions, the difference between
particles and QP's can be ignored. This property is a consequence of
the Ward identities that result from particle-number conservation 
(see Ref.~\onlinecite{Metzner98} for a detailed
discussion). 

Eqs.~(\ref{Phi30},\ref{Phi3}) determine $\chi^{-1}$. If, for
simplicity, we consider the case where only the Landau parameters
$F_0^s$ and $F_0^a$ are non-zero, the matrix $\chi^{-1}$ can easily be
inverted, and we obtain the well-known expressions\cite{Abrikosov63}
\begin{eqnarray}
\chi_{\rm ch}(\tilde q)&=&2N(0)\frac{\Omega_{00}(i\Omega/v_F^*q)}{1+F_0^s
\Omega_{00}(i\Omega/v_F^*q)},\nonumber \\
\chi_{\rm sp}(\tilde q)&=&2N(0)\frac{\Omega_{00}(i\Omega/v_F^*q)}{1+F_0^a
\Omega_{00}(i\Omega/v_F^*q)},
\end{eqnarray}
where 
\begin{equation}
\Omega_{00}(x)=\int_{-1}^{1} \frac{du}{2}\frac{u}{u-x}
= 1-\frac{x}{2}\ln \frac{x+1}{x-1}.
\end{equation}

\section{Renormalization-group approach}
\label{sec:RG}

Recently, RG techniques based on low-energy fermion
effective actions have been applied to interacting fermions in
dimension $d\geq 2$ by many authors (see 
Refs.~\onlinecite{Chitov95,Dupuis96,Chitov98,Dupuis98,Metzner98,Benfatto90,Shankar91,Polchinski92,Shankar94,Nayak94}
and references therein). The finite-temperature RG
approach\cite{Bourbon91} was first applied to Fermi liquids 
in dimension $d\geq 2$ by Chitov and S\'en\'echal.\cite{Chitov95} Contrary to 
other works on the subject, it revealed that the effective interaction 
function in the Landau or zero-sound channel (particle-hole pairs
at small total momentum and energy) does not stay marginal under the RG 
transformation, since its $\beta $-function is not identically zero. From the
RG equations, the standard FLT results have been recovered. 
\cite{Chitov95,Dupuis96,Chitov98,Dupuis98} It has been pointed
out that the bare interaction function of the low-energy fermion effective 
action cannot be identified with the Landau $f$ function.\cite{Dupuis96} 
The latter, along with other observable parameters of a Fermi liquid, is 
obtained at the fixed-point of the RG equations, i.e., when all degrees of
freedom have been integrated out. \cite{Dupuis96,Chitov98,Dupuis98}
The finite temperature RG approach has also given results 
that cannot be obtained within the standard derivation of FLT. 
Chitov and S\'en\'echal, taking into account the interferences between the 
zero-sound (ZS) and exchange (ZS') channels, have obtained RG equations that
satisfy the Pauli principle contrary to the standard microscopic derivation of
FLT. \cite{Chitov98} Performing a two-loop order calculation, the
present author has obtained a  non trivial expression of the
wave-function renormalization factor,\cite{Dupuis98} which was also
obtained from 2D bosonization  \cite{Kwon95} and Ward
Identities. \cite{Metzner98}  

In this section, we show how the response function 
$\chi$ introduced in section \ref{sec:md}, and therefore the
Landau $f$ function, can be calculated using a
a finite-temperature (Kadanoff-Wilson) RG approach. We follow
the procedure used in Ref.~\onlinecite{Dupuis98} to obtain the
compressibility of a Fermi liquid.  
For simplicity we consider a 2D fermion gas.

We write the partition function $Z[h]$ as a functional integral over
Grassmann variables [Eq.~(\ref{Z1})], where, assuming that the high-energy 
degrees of freedom have been integrated out (in a functional sense),
the action describes fermionic degrees of freedom with
$|k-k_F|<\Lambda_0\ll k_F$. We write this low-energy effective action
as  $S^{\Lambda_0}+S^{\Lambda_0}_h$, where 
\begin{eqnarray}
S^{\Lambda_0} &=& -
\sum _{\tilde k,\sigma } \bar\psi ^*_\sigma (\tilde k)(i\omega -\epsilon
_{\bf k} + \mu ) \bar\psi _\sigma (\tilde k)
\nonumber \\ & &
+{1 \over {4\beta \nu }} \sum _{\tilde k_1...\tilde k_4} \sum _{\sigma
_1... \sigma _4} \Gamma ^{\Lambda _0}
_{\sigma _1\sigma _2,\sigma _3\sigma _4} (\tilde k_1,\tilde
k_2,\tilde k_3, \tilde k_4)
\nonumber \\ & & \times  
\bar\psi ^*_{\sigma _4}(\tilde k_4)
\bar\psi ^*_{\sigma _3}(\tilde k_3)
\bar\psi _{\sigma _2}(\tilde k_2)
\bar\psi _{\sigma _1}(\tilde k_1) \nonumber \\ & & \times
\delta _{{\bf k}_1+{\bf k}_2,{\bf k}_3+{\bf k}_4}
\delta _{\omega _1+\omega _2,\omega _3+\omega _4} .
\label{action}
\end{eqnarray} 
It will be shown below that the RG procedure ensures that the action is
always expressed as a function of the QP field $\bar\psi^{(*)}$. In
(\ref{action}), the wave-vectors ${\bf k}$ satisfy $\vert k-k_F \vert
< \Lambda _0$.  The single-particle 
excitations are linearized around the Fermi surface: $\epsilon _{\bf k}-\mu=
v_F(k-k_F)$ where $v_F$ is the bare Fermi velocity. The bare 
(antisymmetrized) two-particle vertex $\Gamma ^{\Lambda_0}$ is assumed to be a
non-singular function of its arguments. The summation over wave vectors is 
defined by
\begin{equation}
{1 \over \nu } \sum _{\bf k}=\int {{d^2{\bf k}} \over {(2\pi )^2}} 
\equiv k_F \int _{k_F-\Lambda _0}^{k_F+\Lambda _0} {{dk} \over {2\pi }} 
\int _0^{2\pi } {{d\theta } \over {2\pi }} ,
\end{equation}
ignoring irrelevant terms at tree-level.

As shown in section \ref{subsec:dffs}, the Landau's functional
$\delta\Phi[\delta n]$ can also be
expressed in terms of the Fermi surface displacements $u_\sigma({\bf
\hat k})\equiv u_\sigma(\theta)$, or equivalently the `density' variations
$\delta\rho_\sigma(\theta)=(k_F/2\pi)u_\sigma(\theta)$. [Here and in
the following, we denote by $\theta$ the direction of a given momentum
$\bf k$, i.e., $\bf \hat k=(\cos\theta,\sin\theta)$.] For purely
technical convenience, we therefore consider an external field
$h_\sigma(\theta)$ that couples directly to the QP density operator
$\hat\rho_\sigma(\theta,\tau)$ and write the action $S_h^{\Lambda_0}$
as
\begin{equation}
S_h^{\Lambda_0}=\sum_\sigma \int \frac{d\theta}{2\pi}h_\sigma(\theta)
\int_0^\beta d\tau \hat \rho_\sigma(\theta,\tau),
\label{action1}
\end{equation}
where 
\begin{eqnarray}
\int_0^\beta d\tau 
\hat \rho_\sigma (\theta,\tau)&=&v_FN(0)
\sum _\omega \int _{k_F-\Lambda_0}^{k_F+\Lambda_0}dk \nonumber \\ && \times 
\lim _{q \to 0} \Bigl [\bar\psi ^*_\sigma 
(\tilde k+\tilde q)\bar\psi _\sigma (\tilde k) \Bigr |_{\Omega=0} \Bigr ].
\end{eqnarray}
We take the $q$-limit as discussed in section \ref{sec:md}. 
$N(0)=k_F/2\pi v_F$ is the bare 2D density of states per spin. 

It is straightforward to show that the microscopic definition of the 
Landau $f$ function [Eqs.~(\ref{chidef1},\ref{chidef2})] becomes:
\begin{eqnarray}
\frac{\nu}{N(0)}[\delta_{\sigma,\sigma'}2\pi\delta(\theta)+N(0)
f_{\sigma\sigma'} (\theta)]=\chi^{*-1}_{\sigma\sigma'}(\theta), && 
\nonumber \\ 
\chi^*_{\sigma\sigma'}(\theta-\theta')=4\pi^2\frac{\beta^{-1}\delta
^{(2)}\ln Z[h]}{\delta h_\sigma(\theta)\delta h_{\sigma'}(\theta')}
\Biggl |_{h=0} . &&
\label{chidef3}
\end{eqnarray}
The meaning of the notation $\chi^*$ is discussed below.

The constraint to have all momenta in the shell $|k-k_F|<\Lambda _0$
restricts the 
allowed scatterings to diffusion of particle-hole, or particle-particle, pairs
with small total momentum ($q<\Lambda _0$). \cite{Shankar94}
Consequently, only two vertex
functions have to be considered: the forward-scattering vertex function and
the BCS vertex function. In the absence of BCS instability, we can neglect 
the latter. We denote by $\Gamma _{\sigma_i}^{\Lambda_0}(\tilde k_1,\tilde k_2;
\tilde q)$ the forward-scattering vertex. [We use the notation 
$\Gamma^{\Lambda_0}_{\sigma _i}=\Gamma^{\Lambda_0}_{\sigma _1\sigma_2,\sigma_3
\sigma_4}$.] $\tilde q=({\bf q},i\Omega)$ where $\bf q$ is the momentum 
transfer ($q<\Lambda_0$) and the bosonic Matsubara frequency $\Omega$ the 
energy transfer. $(\tilde k_1,
\sigma_1)$ and $(\tilde k_2,\sigma_2)$ correspond to the two incoming 
particles, and $(\tilde k_1+\tilde q,\sigma_4)$ and $(\tilde k_2-\tilde q,
\sigma_3)$ to the two outgoing particles. Neglecting the irrelevant dependence
on $k_{1,2}$ and $\omega _{1,2}$, we consider only the coupling function 
$\Gamma ^{\Lambda_0}_{\sigma_i}({\bf \hat k}_1,{\bf \hat k}_2;\tilde
q)\equiv \Gamma ^{\Lambda_0}_{\sigma_i}(\theta_1,\theta_2;\tilde q)$.

The Kadanoff-Wilson RG procedure consists in successive partial integrations of
the fermion field degrees of freedom in the infinitesimal momentum shell
$\Lambda _0e^{-dt}\leq \vert k-k_F\vert \leq \Lambda _0$ where $dt$ is the 
RG generator and $\Lambda (t)=\Lambda
_0e^{-t}$ the effective momentum cut-off at step $t$ . Each
partial integration is followed by a rescaling of radial momenta,
frequencies and fields
(i.e., $\omega '=s\omega $, $k'-k_F=s(k-k_F)$ and $\bar\psi '   
=\bar\psi$ with 
$s=e^{dt}$) in order to let the quadratic part of the action (\ref{action}) 
invariant and to restore the initial value of the cut-off.
The partial integration modifies the parameters of the action which become 
functions of the flow parameter $t$. In the following, we note $\Gamma$
the running (i.e., cut-off dependent) vertex (we do not write its 
$t$-dependence explicitly).
For the purpose of our calculation, it is sufficient to consider
the $q$- and $\Omega$-limits of the forward-scattering vertex:
\begin{eqnarray}
\Gamma _{\sigma _i}^q(\theta _1-\theta _2) &=& \lim _{q\rightarrow 0} \Bigl 
\lbrack \Gamma _{\sigma _i}(\theta _1,\theta _2;\tilde q) 
\Bigl \vert _{\Omega =0} \Bigr \rbrack , \nonumber \\
\Gamma _{\sigma _i}^\Omega (\theta _1-\theta _2) &=& \lim _{\Omega 
\rightarrow 0} \Bigl  \lbrack \Gamma _{\sigma _i}(\theta _1,\theta _2;\tilde
q) \Bigl \vert _{q=0} \Bigr \rbrack . 
\end{eqnarray}

The RG process also generates corrections to the source field $h_\sigma 
(\theta)$ along with higher-order terms in $h$. At step $t$, 
the source term in the action can be written as\cite{note3} (ignoring terms 
of order $h^3$)
\begin{eqnarray}
S_h^\Lambda &=& \sum _{\sigma,\sigma'} \int \frac{d\theta}{2\pi} 
h_\sigma (\theta)
\int \frac{d\theta'}{2\pi} z^{(h)}_{\sigma\sigma'}(\theta-\theta')\int_0^\beta
d\tau\, \hat \rho _{\sigma'}(\theta',\tau) \nonumber \\ &&
-\frac{1}{2} \sum _{\sigma,\sigma'} \int \frac{d\theta}{2\pi}
\frac{d\theta'}{2\pi} h_\sigma (\theta)\beta\chi _{\sigma\sigma'}
(\theta-\theta')h_{\sigma'}(\theta').
\label{Sh}
\end{eqnarray}
We do not write explicitly the dependence of $z^{(h)}$ and $\chi$ on the flow
parameter $t$. 

Note that $z^{(h)}$ and $\chi$ are not physical observables,
since they do not result from the integration over all degrees of freedom. 
Only their fixed-point values $z^{(h)*}$ and $\chi^*$, obtained when 
$\Lambda(t)=0$, are physical (observable) quantities. The Landau $f$
function is therefore related to $\chi^*$. [Note that $\chi^*$
corresponds to what we denoted by $\chi$ in the preceding sections.]

\subsection{one-loop order}

The integration of high-energy degrees of freedom
($|k-k_F|>\Lambda_0$) leading to 
the low-energy effective action (\ref{action}) will in general generate a
wave-function renormalization factor $z_{\Lambda_0}<1$ and an external
field renormalization $z^{(h)}|_{\Lambda_0}$. We ignore these
complications which  
will be discussed in section \ref{subsec:bol}. Thus, the initial conditions of
the RG equations, besides $\Gamma |_{\Lambda_0}=\Gamma^{\Lambda_0}$, are
\begin{eqnarray}
z^{(h)}_{\sigma\sigma'}(\theta)|_{\Lambda_0}&=& 
\delta _{\sigma,\sigma'}2\pi \delta (\theta), \nonumber \\ 
\chi _{\sigma\sigma'}(\theta)|_{\Lambda_0}&=&0.
\end{eqnarray}
The latter equation follows from the condition $T\ll \Lambda_0$ and the
fact that the external field $h$ creates excitations only in the
vicinity of the Fermi surface. 

The external field renormalization at one-loop order is given by (see 
Fig.~2a in Ref.~\onlinecite{Dupuis98})
\begin{eqnarray}
dz^{(h)}_{\sigma\sigma'}(\theta-\theta')&=& \sum_{\sigma''}\int 
\frac{d\theta''}{2\pi}z^{(h)}_{\sigma\sigma''}(\theta-\theta'')
\Gamma ^q_{\sigma''\sigma',\sigma'\sigma''}(\theta''-\theta') \nonumber \\
&& \times \frac{k_F}{2\pi}\int^{'} dk'' 
\frac{1}{\beta}\sum _{\omega ''} [\bar G(\tilde k'')]^2,
\label{zh1}
\end{eqnarray}
where $\bar G(\tilde k)=(i\omega -v_F(k-k_F))^{-1}$ is the QP
propagator. $\int^{'}$ indicates that the  
integration is restricted to the degrees of freedom that are in the 
infinitesimal momentum shell to be integrated out. Using 
\begin{equation}
\frac{1}{\beta}\sum _{\omega ''} \int^{'} dk''
[\bar G(\tilde k'')]^2 = -\frac{\beta _R}{v_F\cosh ^2\beta _R}dt,
\label{sG2}
\end{equation}
we obtain
\begin{equation}
\frac{dz^{(h)}_{\sigma\sigma'}(l)}{dt}= 
-\frac{N(0)\beta _R}{\cosh ^2\beta _R}
\sum _{\sigma''} z^{(h)}_{\sigma\sigma''}(l) 
\Gamma ^q_{\sigma''\sigma',\sigma'\sigma''}(l).
\label{zh2}
\end{equation}
We have introduced the dimensionless inverse temperature $\beta _R=v_F\beta 
\Lambda (t)/2$, and expanded in circular harmonics the quantities appearing in 
(\ref{zh1}), $z^{(h)}_{\sigma\sigma'}(\theta)=\sum _l z^{(h)}_{\sigma\sigma'}
(l)e^{il\theta}$... Eq.~(\ref{zh2}) is solved by introducing 
\begin{eqnarray}
z_+^{(h)}(l)&=&z_{\uparrow\uparrow}^{(h)}(l)+z_{\uparrow\downarrow}^{(h)}(l),
\nonumber \\ 
z_-^{(h)}(l)&=&z_{\uparrow\uparrow}^{(h)}(l)-z_{\uparrow\downarrow}^{(h)}(l),
\end{eqnarray}
and the spin symmetric  ($A^q$) and antisymmetric ($B^q$) parts 
of the two-particle vertex defined by 
\begin{equation}
2N(0)\Gamma _{\sigma _i}^q(l)=A^q_l\delta _{\sigma_1,\sigma_4}
\delta_{\sigma_2,\sigma_3}+B^q_l\mbox {\boldmath $\tau $}
_{\sigma_1\sigma_4} \cdot \mbox {\boldmath $\tau $}_{\sigma_2\sigma_3},
\end{equation}
where $\mbox {\boldmath $\tau $}$ denotes the Pauli matrices. Eq.~(\ref{zh2})
decouples into two independent equations:
\begin{eqnarray}
\frac{d\ln z_+^{(h)}(l)}{dt}&=&-\frac{\beta_R}{\cosh ^2\beta _R}A_l^q, 
\nonumber \\
\frac{d\ln z_-^{(h)}(l)}{dt}&=&-\frac{\beta_R}{\cosh ^2\beta _R}B_l^q.
\label{zh3}
\end{eqnarray}
Eqs.~(\ref{zh3}) have to be supplemented with the one-loop RG equations for 
the vertex functions $A^q$ and $B^q$. As shown in 
Ref.~\onlinecite{Dupuis96}, the latter can be written as
\begin{eqnarray}
{{dA^q_l} \over {dt}} &=& -{{\beta _R} \over {\cosh ^2\beta _R}} {A^q_l}^2
+{{dA^\Omega _l} \over {dt}} , \nonumber \\
{{dB^q_l} \over {dt}} &=& -{{\beta _R} \over {\cosh ^2\beta _R}} {B^q_l}^2
+{{dB^\Omega _l} \over {dt}} .
\label{ABeq}
\end{eqnarray}
The first terms of the rhs of Eqs.~(\ref{ABeq}) are the contribution of the 
ZS graph to the renormalization of $\Gamma^q$. The contribution of the ZS' and 
BCS graphs is taken into account via the second terms on the rhs of 
(\ref{ABeq}) (see Fig.~1 in Ref.~\onlinecite{Dupuis98}).
Because the thermal factor $\beta_R/\cosh ^2\beta _R$ is a strongly peaked 
function of $\Lambda (t)$ near $\Lambda (t)=0$ when $T\to 0$, 
these equations have the approximate solutions
\cite{Dupuis96,Chitov98,Dupuis98}
\begin{eqnarray}
A^q_l(\tau ) &=& {{{A^\Omega _l}^*} \over {1+(1-\tau ){A^\Omega _l}^*}},
\nonumber \\
B^q_l(\tau ) &=& {{{B^\Omega _l}^*} \over {1+(1-\tau ){B^\Omega _l}^*}},
\label{ABflow}
\end{eqnarray}
for $\Lambda (t)\lesssim T/v_F$. ${A^\Omega _l}^*=A^\Omega _l|_{\Lambda(t)=0}$ 
and ${B^\Omega _l}^*=B^\Omega _l|_{\Lambda(t)=0}$ are the fixed-point values 
of $A^\Omega _l$ and $B^\Omega _l$. We have introduced the parameter $\tau =
\tanh \beta_R$ and used $\tanh (\beta v_F\Lambda_0/2)\simeq 1$ for $T\ll 
\Lambda_0/v_F$. Since Eqs.~(\ref{ABflow}) hold beyond one-loop order, we
postpone their detailed derivation to section \ref{subsec:bol}. 
Eqs.~(\ref{zh3}) show that the RG flow of $z^{(h)}$ becomes
significant only at small energy when $\Lambda (t)\lesssim T/v_F$. This allows 
to insert (\ref{ABflow}) into (\ref{zh3}) and obtain 
\begin{eqnarray}
z_+^{(h)}(l) &=& \frac{1}{1+(1-\tau )A_l^{\Omega *}}, \nonumber \\
z_-^{(h)}(l) &=& \frac{1}{1+(1-\tau )B_l^{\Omega *}},
\label{zh4}
\end{eqnarray}
using the initial conditions $z_+^{(h)}(l)|_{\Lambda_0}=z_-^{(h)}(l)|
_{\Lambda_0}=1$. 

The renormalization of $\chi$ is given by (see Fig.~2b in
Ref.~\onlinecite{Dupuis98})
\begin{eqnarray}
d\chi_{\sigma\sigma'}(\theta-\theta') &=& -\sum _{\sigma''}
\int \frac{d\theta''}{2\pi}z^{(h)}_{\sigma\sigma''}(\theta-\theta'')
z^{(h)}_{\sigma'\sigma''}(\theta'-\theta'') \nonumber \\
&& \times \frac{k_F}{2\pi\beta\nu}\sum _{\omega ''} 
\int^{'} dk'' [\bar G(\tilde k'')]^2 .
\label{chi2}
\end{eqnarray}
Expanding (\ref{chi2}) in circular harmonics and using (\ref{sG2}), we obtain
\begin{equation}
\frac{d\chi _{\sigma\sigma'}(l)}{d\tau}=-\frac{N(0)}{\nu}
\sum _{\sigma ''}z^{(h)}_{\sigma\sigma''}(l)z^{(h)}_{\sigma'\sigma''}(l).
\label{chi3}
\end{equation}
To derive (\ref{chi3}) from (\ref{chi2}), we have used the fact that 
$z^{(h)}_{\sigma\sigma'}(\theta)$ is an even function of $\theta$. Thus, we
obtain
\begin{eqnarray}
\frac{d\chi _{\uparrow\uparrow}(l)}{d\tau}&=&-\frac{N(0)}{2\nu}
[z^{(h)}_+(l)^2+z^{(h)}_-(l)^2], \nonumber \\
\frac{d\chi _{\uparrow\downarrow}(l)}{d\tau}&=&-\frac{N(0)}{2\nu}
[z^{(h)}_+(l)^2-z^{(h)}_-(l)^2].
\label{chi4}
\end{eqnarray}
The integration of (\ref{chi4}) using (\ref{zh4}) and the initial condition 
$\chi _{\sigma\sigma'}(l) |_{\Lambda_0}=0$ yields the fixed-point matrix
\begin{eqnarray}
\chi ^*_{\uparrow\uparrow}(l)&=&\frac{N(0)}{2\nu}\Biggl [ 
\frac{1}{1+A^{\Omega *}_l}+\frac{1}{1+B^{\Omega *}_l} \Biggr ],
\nonumber \\
\chi ^*_{\uparrow\downarrow}(l)&=&\frac{N(0)}{2\nu}\Biggl [ 
\frac{1}{1+A^{\Omega *}_l}-\frac{1}{1+B^{\Omega *}_l} \Biggr ].
\end{eqnarray}
The inverse matrix is 
\begin{eqnarray}
\chi ^{*-1}_{\uparrow\uparrow}(l)&=&\frac{\nu}{2N(0)}(
2+A^{\Omega *}_l+B^{\Omega *}_l), \nonumber \\
\chi ^{*-1}_{\uparrow\downarrow}(l)&=&\frac{\nu}{2N(0)}(
A^{\Omega *}_l-B^{\Omega *}_l).
\label{Cinv}
\end{eqnarray}
 
From the microscopic definition of the Landau $f$ function
[Eq.~(\ref{chidef3})], we deduce
\begin{equation}
f_{\sigma\sigma'}(\theta)=\Gamma ^{\Omega*}_{\sigma\sigma',\sigma'\sigma}
(\theta),
\label{fL}
\end{equation}
a result fist obtained in Ref.~\onlinecite{Dupuis96}.

\subsection{Beyond one-loop}
\label{subsec:bol}

In this section, we show that higher-order loop contributions do not change
Eq.~(\ref{fL}), except for a dependence of $f$ on the wave-function 
renormalization factor. The RG equations can be solved exactly if we
make the three following assumptions (see also Ref.~\onlinecite{Dupuis98}, 
section 5): we assume the existence of well-defined
QP's in the vicinity of the Fermi surface (i). Except the one-loop ZS graph,
all the graphs are regular in the limit $\tilde q\to 0$ (i.e., give the
same contribution in the $q$- and $\Omega$-limits) (ii), and give a smooth 
contribution to the RG flow of various physical quantities (iii). 
[Because of the thermal factor $\beta /4\cosh^2\beta_R$, 
which is a strongly peaked function of $\Lambda (t)$ around $\Lambda (t)=0$ for
$T\to 0$, the ZS graph yields a singular contribution (with respect to $\Lambda
(t)$) to the RG flow of $\Gamma ^q$.] Assumption (ii) can be explicitly 
verified at one-loop order. Assumption (iii) is only `approximate' in a sense
that is further discussed below. 

Let us first consider the RG equations for the two-particle vertex. 
They can be written as\cite{Dupuis96}
\begin{eqnarray}
{{dA^q_l} \over {dt}} &=&
{{dA^q_l} \over {dt}} \Biggl \vert _{\rm ZS}+
{{dA^\Omega_l} \over {dt}}, \label{rgv1} \\ 
{{dA^q_l} \over {dt}} \Biggl \vert _{\rm ZS}&=&
- {{\beta _R} \over {\cosh ^2(\beta _R)}}{A^q_l}^2,
\label{rgv2}
\end{eqnarray}
and a similar equation for $B^q$. We have written explicitly the contribution
of the one-loop ZS graph which distinguishes between the $q$- and 
$\Omega$-limits. The second term on the rhs of (\ref{rgv1}) includes the 
contribution of other one-loop graphs (ZS' and BCS channels) 
as well as higher-order loop corrections. [Here we use assumption (ii) and
the fact that the ZS graph vanishes in the $\Omega$-limit.] 

It is tempting to neglect the term $dA_l^\Omega/dt$ in (\ref{rgv1})
(see for instance Ref.~\onlinecite{Shankar94}).
Because of phase-space restrictions (which result from momentum 
conservation at the  interaction vertices), the corresponding diagrams are
suppressed by the small parameter $\Lambda_0/k_F$ if one considers only
low-energy states $|k-k_F|\leq \Lambda_0$ (assuming that high-energy states 
($|k-k_F|>\Lambda_0$) have already been integrated out). This property results
from a frustration of the interferences between channels in dimension $d\geq 
2$. \cite{note5} Solving (\ref{rgv1}) without the last term of the rhs is 
equivalent to an RPA calculation in the ZS channel.\cite{Chitov95,Chitov98}
This RPA calculation can also be done using standard diagrammatic theory.
Within this approximation, the
Landau $f$ function is naturally identified with the bare vertex function 
$\Gamma^{\Lambda_0}|_{\tilde q=0}$. [We do not need to distinguish between 
the $q$- and the $\Omega$-limits, since $\Gamma^{\Lambda_0}$ is a regular 
function of its arguments.] A major drawback of this approximation is that 
the momentum scale $\Lambda_0$, which is not a physical scale in the problem, 
enters the definition of the Landau parameters in an essential way. [For a 
further discussion of this RPA approximation, see Ref.~\onlinecite{Chitov98}.] 

In Refs.~\onlinecite{Dupuis96,Dupuis98}, the present author and
G. Chitov have proposed another approach 
to solve (\ref{rgv1},\ref{rgv2}), which bears some similarities with Landau's 
solution of the Bethe-Salpeter equation for the two-particle
vertex. Besides the  
fact that it is not suppressed by the small parameter $\Lambda_0/k_F$, the ZS 
graph presents another interesting feature. Its contribution to the RG flow of 
$A^q$ and $B^q$ becomes singular (with respect to $\Lambda(t)$) at low 
temperature since $\beta_R/4\cosh^2\beta_R$ is exponentially suppressed for
$\Lambda(t)\gtrsim T$. To see how this allows to solve 
Eqs.~(\ref{rgv1},\ref{rgv2}), we integrate the latter (using $A^q|_{\Lambda_0}
=A^\Omega|_{\Lambda_0}$):
\begin{equation}
A^q_l(t)=A^\Omega_l(t)-\int_0^t dt' \frac{\beta_R}{\cosh^2\beta_R}
A^q_l(t')^2.
\label{A1}
\end{equation}
Iterating (\ref{A1}), we obtain
\begin{equation}
A^q_l(t)=A^\Omega_l(t)-\int_0^t dt' \frac{\beta_R}{\cosh^2\beta_R}
A^\Omega_l(t')^2+\cdots 
\label{A2}
\end{equation}
According to assumption (iii), $A_l^\Omega(t')$ is a smooth function
of the cut-off $\Lambda(t')$ when $\Lambda(t')\lesssim T/v_F$. This implies
$A_l^\Omega(t')|_{\Lambda(t')\lesssim T/v_F}\simeq A_l^{\Omega*}$ at low 
temperature. Since, on the other hand, $\beta_R/4\cosh^2\beta_R$ is strongly 
peaked for $\Lambda(t')\lesssim T/v_F$, we can replace $A_l^\Omega(t')$ in 
the rhs of (\ref{A2}) by its fixed-point value $A_l^{\Omega*}$.
The RG equation for $A^q$ then becomes (for $\Lambda(t)\lesssim T/v_F$)
\begin{equation}
A^q_l(t)=A^{\Omega*}_l-\int_0^t dt' \frac{\beta_R}{\cosh^2\beta_R}
A^q_l(t')^2.
\label{A3}
\end{equation}
Eq.~(\ref{A3}) corresponds to a decoupling of 
the ZS channel from the other channels: when $\Lambda(t)\gtrsim T/v_F$, the
ZS graph does not contribute to the RG flow, and both $A^q$ and 
$A^\Omega$ evolve smoothly when the cut-off decreases (with $A^q\simeq 
A^\Omega$); when $\Lambda(t)$ varies from $\sim T/v_F$ to $0$, 
only the ZS graph gives a significant contribution (assumption (iii)) and 
drives $A^q$ towards its fixed-point value $A^{q*}$ (but does not 
contribute to the renormalization of $A^\Omega$). The solution of 
Eq.~(\ref{A3}) and the analog equation for $B^q$ is given 
by (\ref{ABflow}). We now obtain an RPA-like relation between two {\it
fixed-point quantities}:
$A^{q*}_l=A^{\Omega*}_l/(1+A^{\Omega*}_l)$. Both quantities are
physical in the sense that they are obtained by integrating all the
degrees of freedom.

One should combine the RG equations of $A^q$ and $B^q$ with RG equations of
the Fermi velocity $v_F(t)$ and the wave-function renormalization factor 
$z(t)$. Since the ZS loop does not appear in self-energy corrections, $v_F(t)$ 
and $z(t)$ are smooth functions of the cut-off $\Lambda(t)$. Thus, the 
argument leading to (\ref{A3}) and (\ref{ABflow}) also holds when self-energy
corrections are taken into account. When solving (\ref{A3}), one should replace
$v_F(t)$ and $z(t)$ by their fixed-point values $v_F^*$ and $z^*$.
\cite{Dupuis98}

If assumption (iii) were to hold exactly, Eqs.~(\ref{ABflow}) would be exact
for $\Lambda(t)\lesssim T/v_F$ and $T\to 0$. It has been pointed out 
in Ref.~\onlinecite{Chitov95} that the ZS' graph also gives a singular 
contribution to the flow of $\Gamma^q(\theta-\theta')$ when the two incoming 
particles have parallel momenta ($\theta-\theta'\to 0$). 
In this case, the ZS' loop becomes identical to the ZS loop (see
Fig.~1 of Ref.~\onlinecite{Dupuis98}). Thus, assumption (iii) is not 
exact and Eqs.~(\ref{ABflow}) hold only approximately. The singularity in the 
ZS' channel is however restricted to very small angles $|\theta-\theta'|
\lesssim T/v_Fk_F$. Only for those angles do the ZS and ZS' channels interfere
when $\Lambda(t)\lesssim T/v_F$. Consequently, only the components 
$\Gamma^q(l)$ with $l\gtrsim v_Fk_F/T$ are affected by the small-angle 
singularity in the ZS' channel. For most physical quantities (specific
heat, effective mass, compressibility...), Eqs.~(\ref{ABflow}) remain an
excellent approximation. [The 
singularity of the ZS' graph becomes crucial if one is
precisely interested in the value of $\Gamma^q(\theta-\theta')$ for $\theta-
\theta'\to 0$.\cite{Chitov98}] This conclusion has been checked by explicit
calculation at one-loop order by Chitov and S\'en\'echal. \cite{Chitov98} 

At each step of the RG transformation, the field is rescaled according to
$\bar\psi'=[z(dt)]^{-1/2}\bar\psi$, where $z(dt)$ is the contribution to the 
wave-function renormalization factor $z(t)$ when the flow parameter $t$ 
increases by $dt$: $z(t+dt)=z(t)z(dt)$. This ensures that the
propagator keeps the form $(i\omega -v_F(t)k)^{-1}$.\cite{Dupuis98} Thus the
field $\bar\psi$ refers to QP's as anticipated. 

Now we consider the renormalization of the source field. Although the
action is expressed only in terms of the QP field, $h$ (as any other
external field) couples a priori to the bare fermions. The coupling
between the external field and the incoherent part of the single-particle
spectral function appears indirectly through some renormalization of
the external field (see for instance
Ref.~\onlinecite{Shankar94} for a discussion). Here we want to eliminate such
renormalizations since $h$ couples directly to the QP's. To distinguish
between the coherent (i.e., due to the QP's) and incoherent responses
to the field $h$, we also consider the response to the field
$h^\Omega$ that couples to the incoherent part of the density:
\begin{eqnarray}
\int_0^\beta d\tau 
\hat \rho^\Omega_\sigma (\theta,\tau)&=&v_FN(0)
\sum _\omega \int _{k_F-\Lambda_0}^{k_F+\Lambda_0}dk \nonumber \\ && \times 
\lim _{\Omega \to 0} \Bigl [\bar\psi ^*_\sigma 
(\tilde k+\tilde q)\bar\psi _\sigma (\tilde k) \Bigr |_{q=0} \Bigr ] .
\label{hOm}
\end{eqnarray}
When $T\to 0$, $h^\Omega$ cannot create coherent particle-hole pairs (i.e.,
quasi-particle-quasi-hole pairs) since the $\Omega$-limit is taken in
(\ref{hOm}).\cite{note4} On the contrary, $h$ couples to 
both the coherent and incoherent parts.
$n$-loop contributions ($n\geq 2$) to the field 
renormalization factor $z^{(h)}$ do not distinguish between $h$ and $h^\Omega$.
The reason is that the singular ZS loop does not appear in $n$-loop ($n\geq 2$)
diagrams. Therefore, the latter correspond to coupling of the external field 
to the incoherent part.
On the contrary, the one-loop graph considered in the preceding section 
vanishes in the $\Omega$-limit and corresponds to coupling of the external
field to the coherent part. Since we want $h$ to couple only to the QP's, 
only the one-loop diagram for $z^{(h)}$ has to be taken into account. 
QP's are obtained not only by filtering out the incoherent part of the
propagator, but also by rescaling the field according to $\bar\psi'
=[z(dt)]^{-1/2}\bar\psi$. The latter implies
a concomitant rescaling of the field $h'=hz(dt)$ to ensure that
$h\bar\psi^*\bar\psi=h'\bar\psi^{'*}\bar\psi'$. We conclude that the
only renormalization of the external field comes from the one-loop
contribution to $z^{(h)}$ considered in the preceding section.  

We also note that the diagram shown in Fig.~2b of Ref.~\onlinecite{Dupuis98} 
is the only one contributing to the susceptibility in the
Kadanoff-Wilson scheme, since it is the only diagram of order $O(h^2)$
(and $O(dt)$) generated by the RG procedure. Thus, we obtain the same
RG equations as in the one-loop calculation, the bare Fermi velocity being
replaced by its fixed-point value $v_F^*$. Eq.~(\ref{fL}) holds at all order
in a loop expansion (when the small-angle singularity in the ZS' channel is 
neglected). Since the $\bar\psi$'s have been rescaled at each step of the
RG transformation, we eventually come to 
\begin{equation}
f_{\sigma\sigma'}(\theta)=        {z^*}^2
\Gamma ^{\Omega*}_{\sigma\sigma',\sigma'\sigma}(\theta),
\label{Gamfin}
\end{equation}
where $\Gamma ^{\Omega*}_{\sigma_i}$ now refers to the bare fermions, and 
$z^*$ is the fixed-point value of $z(t)$. Eq.~(\ref{Gamfin}) agrees
with the conclusion of Ref.~\onlinecite{Dupuis98}.

\section{Conclusion}

We have proposed a new microscopic description of Fermi liquids, which
extends some early ideas of the statistical FLT. It is based on the
introduction of an effective potential (in the sense of field theory)
$\Phi[n]$, which is obtained from the free energy by a Legendre
transformation. 

In the more general case, the effective potential $\Phi[n]$ is
a functional of the Wigner distribution function $n\equiv \lbrace
n_{{\bf k}\sigma}({\bf r},t)\rbrace$. Small variations
$\delta\Phi[\delta n]$ around the equilibrium value are 
parameterized by the Landau $f$ function, which describes the interaction
between QP's. $f$ has a precise microscopic definition in terms of the
susceptibility $\chi$ introduced in section
\ref{sec:md}. Using the standard assumptions of FLT, we have shown that
this microscopic definition yields the usual identification between
$f$ and the $\Omega$-limit of the forward-scattering two-particle vertex. 
The effective potential $\delta\Phi[\delta n]$ yields
both the response functions of the Fermi liquid and the quantum
Boltzmann equation satisfied by $n_{{\bf k}\sigma}({\bf r},t)$. In the
static and uniform limit, $\delta\Phi[\delta n]$ is nothing but the variation
of the thermodynamic potential corresponding to a change $\delta n$ of
the QP distribution function. $\delta\Phi[\delta n]$ was first
introduced by Landau on phenomenological grounds to describe Fermi
liquids. Thus, the effective potential describes both static and
dynamic properties of Fermi liquids in a unified framework. It should
be noted that this description does not rely on any semiclassical
assumption. 

The explicit calculation of $\delta\Phi[\delta n]$ can be in principle
extended to more complicated and/or realistic situations, for instance
by taking into account the presence of impurities and the effect of an
electric field.

\section*{Acknowledgments}

I would like to thank G. Chitov and A.M.-S. Tremblay for useful
discussions. This work was partially supported by the NSF under Grant
DMR--9417451 and by the Packard Foundation.




\begin{references}

\bibitem[*]{present} Present address.     

\bibitem{Landau57} L.D. Landau, JETP {\bf 3}, 920 (1957); {\bf 5}, 101
(1957).

\bibitem{Landau59}L.D. Landau, JETP {\bf 8}, 70 (1959). 

\bibitem{Abrikosov63} A.A. Abrikosov, L.P. Gor'kov, and I.E. Dzyaloshinski,
{\it Methods of Quantum Field theory in Statistical Physics} (Dover,
New-York, 1963); P. Nozi\`eres, {\it Interacting Fermi Systems} 
(Benjamin, New-York, 1964); G. Baym and C. Pethick, {\it Landau Fermi-liquid
theory} (Wiley, New York, 1991). 

\bibitem{Balian61} R. Balian, C. Bloch, and C. De Dominicis, Nucl. Phys. 
{\bf 25}, 529 (1961);  Nucl. Phys. {\bf 27}, 294 (1961).

\bibitem{Balian64} R. Balian and C. De Dominicis, Physica {\bf 30}, 1927 
(1964); Physica {\bf 30}, 1933 (1964); Ann. Phys. (N.Y.) {\bf 62}, 229 (1971). 

\bibitem{Luttinger68} J.M. Luttinger, Phys. Rev. {\bf 174}, 263 (1968).

\bibitem{Chitov95} G.Y. Chitov and D. S\'en\'echal, Phys. Rev. B. {\bf52},
129 (1995).

\bibitem{Dupuis96} N. Dupuis and G.Y. Chitov, Phys. Rev. B {\bf 54}, 3040
(1996). 
  
\bibitem{Chitov98}  G.Y. Chitov and D. S\'en\'echal, Phys. Rev. B {\bf 57}, 
1444 (1998). 

\bibitem{Dupuis98} N. Dupuis, Eur. Phys. J. B {\bf 3}, 315 (1998).  

\bibitem{note7} The interpretation of $u_\sigma({\bf \hat k})$ as a
Fermi surface displacement comes from the observation that $\delta
n_{{\bf k}\sigma}=\lim_{u_\sigma({\bf \hat k})\to 0} n_F(\tilde
\epsilon _{{\bf k}\sigma }-\mu -v_F^*u_\sigma({\bf \hat
k}))-n_F(\tilde \epsilon _{{\bf k}\sigma }-\mu)$. 

\bibitem{Pomeranchuk59} I.Ia. Pomeranchuk, JETP {\bf 8}, 361 (1959). 

\bibitem{Lebellac} See, for instance, M. Le Bellac, {\it Quantum and
statistical field theory} (Oxford, New York, 1991). 

\bibitem{Negele} See, for instance, J.W. Negele and H. Orland, {\it Quantum 
Many Particle Systems} (Addison-Wesley, New York, 1988). 

\bibitem{BdD} A similar Legendre transformation has been performed by Balian 
{\it et al.} See in particular the second publication of
Ref.~\onlinecite{Balian61}, section 5.  

\bibitem{note4} At $T=0$, it is not possible to create a coherent 
particle-hole pair (i.e., a quasi-particle-quasi-hole pair) with a
finite total energy and a vanishing total momentum  because of the
Pauli principle. 

\bibitem{note6} The two-particle vertex
$\Gamma_{\sigma_1\sigma_2,\sigma_3\sigma_4}(\tilde k_1,\tilde k_2,\tilde
k_3,\tilde k_4)$ is defined such that the direct interaction process
corresponds to $(\tilde k_1,\sigma_1)\to (\tilde k_4,\sigma_4)$ and 
$(\tilde k_2,\sigma_2)\to (\tilde k_3,\sigma_3)$.

\bibitem{Wigner32} E. Wigner, Phys. Rev. {\bf 40}, 749 (1932);
M. Hillery, R.F. O'Connell, M.O. Scully, and E.P. Wigner,
Phys. Rep. {\bf 106}, 121 (1984). 

\bibitem{Mahan} G.D. Mahan, {\it Many-particle physics} (Plenum Press,
New York, 1990). 

\bibitem{Negele1} See also Ref.~\onlinecite{Negele}, chap.~5. 

\bibitem{Metzner98} W. Metzner, C. Castellani, and C. Di Castro,  
Adv. in Phys. {\bf 47}, 317 (1998). 

\bibitem{Benfatto90} G. Benfatto and G. Gallavotti,
Phys. Rev. B {\bf 42}, 9967 (1990); J. Stat. Phys. {\bf 59}, 541 (1990).

\bibitem{Shankar91} R. Shankar, Physica A {\bf 177}, 530 (1991).

\bibitem{Polchinski92} J. Polchinski, in {\it Proceedings of the 1992
Theoretical Advanced Studies Institute in Elementary Particle Physics},
ed. J. Harvey and J. Polchinski (World scientific, Singapore, 1993).

\bibitem{Shankar94} R. Shankar, Rev. Mod. Phys. {\bf 66}, 129 (1994).

\bibitem{Nayak94} C. Nayak and F. Wilczek, Nucl. Phys. B {\bf 430}, 534 (1994);
Int. J. Mod. Phys. B {\bf 10}, 897 (1996).

\bibitem{Bourbon91} The finite-temperature (Kadanoff-Wilson) RG approach to 
fermion systems has been introduced and applied to 1D and quasi-1D conductors
by Bourbonnais and Caron: C. Bourbonnais and L.G Caron, Physica {\bf
143}B, 450 (1986); {\it ibid}, 453. For a review, see C. Bourbonnais
and L. Caron, Int. J. Mod. Phys. B {\bf 5}, 1033 (1991).

\bibitem{Kwon95} See  A. Houghton, H.-J. Kwon, and J. B. Marston, 
cond-mat/9810388 (1998); P. Kopietz, {\it Bosonization
of interacting fermions in arbitrary dimensions} (Springer-Verlag,
1997); and references therein. 

\bibitem{note3} There is also a contribution to the free energy which is
linear in the source field. This contribution does not affect the calculation
of $\chi$ and is not considered here. 

\bibitem{note5} The small parameter $\Lambda_0/k_F$ appears explicitly in the
field theoretical scheme. In the Kadanoff-Wilson scheme, the frustration of
the interference between channels appears in a slightly more indirect way:
see Ref.~\onlinecite{Dupuis98}



\end{references}
\end{document}